%% file: ms.tex
\documentclass[acmsmall,authorversion]{acmart}

\usepackage[utf8]{inputenc}
\usepackage[T1]{fontenc}
\usepackage{mathtools}

\acmJournal{PACMPL}
\acmVolume{1}
\acmNumber{} 
\acmArticle{1}
\acmYear{2020}
\acmMonth{11}
\acmDOI{} 
\startPage{1}

\setcopyright{none}

\providecommand{\tightlist}{%
  \setlength{\itemsep}{0pt}\setlength{\parskip}{0pt}}

\usepackage{microtype}

\newcommand{\autocite}[1]{\cite{#1}}

\usepackage{framed}
\usepackage{booktabs}
\usepackage{longtable}
\usepackage{bcprules} 

\usepackage{ebproof}  
\usepackage{graphicx} 
\usepackage{xfrac}    
\usepackage{color}    
\usepackage[normalem]{ulem} 

\usepackage{listings}
\usepackage{amsmath}
\usepackage{ebproof}
\usepackage{subcaption}
\usepackage{relsize}   
\usepackage{xspace}

\usepackage{cleveref} 
\crefname{subsection}{subsection}{subsections}

\definecolor{light-gray}{gray}{0.95}

\newtheorem*{thm}{Theorem}

\usepackage{enumitem}
\setlist[enumerate]{font=\mdseries}

\lstdefinelanguage{scala}{
  alsoletter={@,=,>},
  morekeywords={abstract, case, class, def,
        else, extends, false, free, if, implicit, match,
        object, true, val, var, while, sealed,
        for, dependent, null, type, with, try, catch, finally,
        import, final, return, new, override, this, trait,
        private, public, protected, package, throw},
  sensitive=true,
  morecomment=[l]{//},
  morecomment=[s]{/*}{*/},
  morestring=[b]",
}
\begin{document}

\title[Coming to Terms with Your Choices: An Existential Take on Dependent Types]{Coming to Terms with Your Choices}
\subtitle{An Existential Take on Dependent Types}


\author{Georg Schmid}
\affiliation{
  \department{LARA}              
  \institution{EPFL}            
  \country{Switzerland}                    
}
\email{georg.schmid@epfl.ch}          

\author{Olivier Blanvillain}
\affiliation{
  \department{LAMP}              
  \institution{EPFL}            
  \country{Switzerland}                    
}
\email{olivier.blanvillain@epfl.ch}          

\author{Jad Hamza}
\affiliation{
  \department{LARA}              
  \institution{EPFL}            
  \country{Switzerland}                    
}
\email{jad.hamza@epfl.ch}          

\author{Viktor Kunčak}
\affiliation{
  \department{LARA}              
  \institution{EPFL}            
  \country{Switzerland}                    
}
\email{viktor.kuncak@epfl.ch}          

\input{rules.tex}

\begin{abstract}
\input{abstract.md}
\end{abstract}




\maketitle

\sloppy
\input{content.tex}


\bibliography{}



\end{document}

%% file: rules.tex
\newcommand\vs{\vspace{4mm}}
\newcommand\nvs{\vspace{-0.7em}}

\newcommand{\oursystem}{$\lambda${\tiny $\mathop{\protect\vphantom{X}}^\text{nd}_{<:\{\}}$}\xspace}
\newcommand{\oursystemlow}{$\lambda${\tiny $\mathop{\protect\vphantom{X}}^\text{det}_{<:\{\}}$}\xspace}

\newcommand\rulename[1]{\textsc{#1}}
\definecolor{ignoreGrey}{gray}{0.5}
\newcommand\grey[1]{\textcolor{ignoreGrey}{#1}}

\newcommand{\defeq}{\vcentcolon=}

\newcommand\abs[3]{\lambda{}#1\!:\!#2.\,#3}   
\newcommand\eabs[2]{\lambda{}#1.\,#2}         
\newcommand\tabs[3]{\Pi{}#1\!:\!#2.\,#3}  
\newcommand\nil{\text{nil}}
\newcommand\cons{\text{cons}}
\newcommand\match{\text{match}}
\newcommand\fix{\text{fix}}
\newcommand\some{\text{some}}
\newcommand\unpack{\text{unpack}}
\newcommand\choos{\text{choose}}
\newcommand\Nil{\text{Nil}}
\newcommand\Cons{\text{Cons}}
\newcommand\Match{\text{Match}}
\newcommand\Fix{\text{Fix}}

\newcommand\letb{\text{let}~}

\newcommand\Any{\text{Top}}
\newcommand\List{\text{List}}
\newcommand\Unit{\text{Unit}}
\newcommand\Nat{\text{Nat}}
\newcommand\Trail{\text{Trail}}

\newcommand\sugar[1]{\langle #1 \rangle}
\newcommand\sugarC[2]{\langle\!\langle #2 \rangle\!\rangle^{#1}}
\newcommand\exsugar[1]{#1}  
\newcommand\refinement[1]{\lbrace\lbrace~v:\Any~|~v \equiv #1 ~\rbrace\rbrace}
\newcommand\refinementOf[3]{\lbrace\lbrace~v:#1~|~#2 \equiv #3 ~\rbrace\rbrace}
\newcommand\existsTerm[2]{{\exists}~#1\!:\!#2.}
\newcommand\singleton[1]{\lbrace #1 \rbrace}
\newcommand\singletonOf[2]{\lbrace #1 \rbrace_{#2}}
\newcommand\texists[3]{\exists#1\!:\!#2.\,#3}
\newcommand\widen[1]{\lceil #1 \rceil}
\newcommand\untangle[1]{\text{untangle}(#1)}
\newcommand\untangl[1]{\mathcal{U}(#1)}
\newcommand\untanglWrap[2]{\mathcal{W}_{\!x}(#1, #2)}
\newcommand\trailsOf[2]{\text{trailsOf}_#1(#2)}
\newcommand\tnorm[2]{#1 \rightarrow_N #2}
\newcommand\listmatch[3]{{#1}~\match~{#2};\,x,y \Rightarrow {#3}}
\newcommand\fixd[4]{\fix_{#1}\!(x\!:\!{#2} \Rightarrow {#3},~{#4})}
\newcommand\tlistmatch[3]{{#1}~\Match~{#2};\,x,y \Rightarrow {#3}}
\newcommand\letin[3]{\text{let}~#1 = #2~\text{in}~#3}
\newcommand\fv[1]{\textit{fv}(#1)}
\newcommand\inhabits[2]{#2 \ni #1}
\newcommand\concatop{\text{+\!\!+}}
\newcommand\traildot[2]{{#1}.{#2}}
\newcommand\FR{System~FR\xspace}

\newcommand{\Hsquare}{\text{\fboxsep=-.2pt\fbox{\rule{0pt}{1ex}\rule{1ex}{0pt}}}}
\newcommand\reducesBD[2]{#1 \rightarrow^*_{\beta\delta} #2}
\newcommand\reducesBDOne[2]{#1 \rightarrow_{\beta\delta} #2}
\newcommand\reducesBoxOne[2]{#1 \rightarrow_{\Hsquare} #2}
\newcommand\reduces[2]{#1 \rightarrow_\beta^*#2}
\newcommand\reducesOne[2]{#1 \rightarrow_\beta#2}
\newcommand\subst[2]{[ #1 \mapsto #2 ]}
\renewcommand\vs{\vspace{3mm}}

\newcommand\todo[1]{\noindent\textcolor{red}{\textbf{TODO:} #1}\\}
\newcommand\todoDone[1]{\todo{\sout{#1}}}

\newcommand\inferenceRules[1]{
  \begin{figure}[!tb]
  \begin{framed}
  \begin{minipage}{0.35\linewidth}
    \infrule[TVar]
      { \Gamma(x) = T }
      { \Gamma \vdash x \Uparrow \singletonOf{x}{T} }
  \end{minipage}
  \begin{minipage}{0.55\linewidth}
    \infrule[TAbs]
      { \Gamma, x : \exsugar{S} \vdash t \Uparrow T }
      { \Gamma \vdash \abs{x}{S}{t} \Uparrow \singletonOf{\abs{x}{\exsugar{S}}{\exsugar{t}}}{\tabs{x}{\exsugar{S}}{T}} }
  \end{minipage}

  \begin{minipage}{0.90\linewidth}
    \vs\infrule[TApp]
      { \Gamma \vdash t_1 \Uparrow V \quad \widen{V} = \tabs{x}{S}{T} \quad \Gamma \vdash t_2 \Downarrow S }
      { \Gamma \vdash t_1~t_2 \Uparrow T \subst{x}{\exsugar{t_2}} }
  \end{minipage}

  \begin{minipage}{0.90\linewidth}
    \vs\infrule[TFix]
      { \Gamma, x : \exsugar{T} \vdash t_1 \Downarrow \exsugar{T} \quad \Gamma \vdash t_2 \Downarrow \exsugar{T} }
      { \Gamma \vdash \fixd{n}{T}{t_1}{t_2} \Uparrow \singletonOf{\exsugar{\fixd{n}{T}{t_1}{t_2}}}{\exsugar{T}} }
  \end{minipage}


  \begin{minipage}{0.90\linewidth}
    \infrule[TNil]
      {~}
      { \Gamma \vdash \nil \Uparrow \singletonOf{\exsugar{\nil}}{\exsugar{\List}} }
  \end{minipage}

  \begin{minipage}{0.90\linewidth}
    \vs\infrule[TCons]
      { \Gamma \vdash t_1 \Uparrow T_1 \quad \Gamma \vdash t_2 \Uparrow T_2 \quad \Gamma \vdash T_2 <: \exsugar{\List} }
      { \Gamma \vdash \cons~t_1~t_2 \Uparrow \singletonOf{\cons~t_1~t_2}{\exsugar{\Cons\,T_1\,T_2}} }
  \end{minipage}

  \begin{minipage}{0.90\linewidth}
    \vs\infrule[TMatch]
      { \Gamma \vdash t_1 \Downarrow \exsugar{\List} \quad \Gamma \vdash t_2 \Uparrow T_2 \quad \Gamma, x: \Any, y: \exsugar{\List} \vdash t_3 \Uparrow T_3 }
      { \Gamma \vdash t_1~\match~t_2;~x,y \rightarrow t_3 \Uparrow \singletonOf{\exsugar{t_1~\match~t_2;~x,y \rightarrow t_3}}{\exsugar{t_1\,\Match\,T_2;\,x,y\,\Rightarrow\,T_3}} }
  \end{minipage}

  \begin{minipage}{0.45\linewidth}
    \infrule[TDot]
      { \Gamma \vdash t \Downarrow \Trail }
      { \Gamma \vdash \traildot{t}{k} \Uparrow \singletonOf{\traildot{t}{k}}{\Trail} }
  \end{minipage}
  \begin{minipage}{0.45\linewidth}
    \vs\infrule[TCheck]
      { \Gamma \vdash t \Uparrow T' \quad \Gamma \vdash T' <: T }
      { \Gamma \vdash t \Downarrow T }
  \end{minipage}
  \vs

  $$
  \begin{aligned}
    \widen{\singletonOf{t}{U}} \defeq{}\ &
      \begin{minipage}{0.65\linewidth}\nvs
        \[
        \hspace{-1.5em}
        \begin{cases}
          \tabs{x}{S}{\singletonOf{t~x}{T}} & \text{if } U = \tabs{x}{S}{T} \text{ for some types } S,T\\
          \widen{U} & \text{otherwise}
        \end{cases}
        \]
      \end{minipage}\\
    \widen{T} \defeq{}\ & T \quad \text{if there exist no $t$ and $U$ such that } T = \singletonOf{t}{U}
  \end{aligned}
  $$
  \end{framed}
  \caption{#1}
  \label{inferenceRules}
  \end{figure}
}

\newcommand\subtypingRulesNew[1]{
  \begin{figure}[!tb]
  \begin{framed}

  \begin{minipage}{0.48\linewidth}
    \nvs\infrule[SubRefl]
      {~}
      { \Gamma \vdash T <: T }
  \end{minipage}
  \begin{minipage}{0.48\linewidth}
    \nvs\infrule[SubTop]
      {~}
      { \Gamma \vdash T <: \Any }
  \end{minipage}

  \begin{minipage}{0.43\linewidth}
    \vs\infrule[SubSing]
      { \Gamma \vdash T_1 <: T_2 }
      { \Gamma \vdash \singletonOf{t}{T_1} <: T_2 }
  \end{minipage}
  \begin{minipage}{0.53\linewidth}
    \vs\infrule[SubExistsLeft]
      { \Gamma, x:S \vdash T <: U }
      { \Gamma \vdash \texists{x}{S}{T} <: U }
  \end{minipage}

  \begin{minipage}{0.96\linewidth}
    \vs\infrule[SubExistsRight]
      { \singletonOf{t}{U} = \text{solve}_x(T_1, S, T_2) \quad
        \Gamma \vdash \singletonOf{t}{U} <: S \quad
        \Gamma \vdash T_1 <: T_2 \subst{x}{t} }
      { \Gamma \vdash T_1 <: \texists{x}{S}{T_2} }
  \end{minipage}

  \begin{minipage}{0.43\linewidth}
    \vs\infrule[SubCons1]
      {~}
      { \Gamma \vdash \exsugar{\Cons~S~T} <: \exsugar{\List} }
  \end{minipage}
  \begin{minipage}{0.53\linewidth}
    \vs\infrule[SubCons2]
      { \Gamma \vdash \exsugar{S_1} <: \exsugar{S_2}  \quad  \Gamma \vdash \exsugar{T_1} <: \exsugar{T_2} }
      { \Gamma \vdash \exsugar{\Cons~S_1~T_1} <: \exsugar{\Cons~S_2~T_2} }
  \end{minipage}

  \begin{minipage}{0.96\linewidth}
    \vs\infrule[SubMatch]
      { \Gamma \vdash S_2 <: T \quad \Gamma, x : \Any, y : \exsugar{\List} \vdash S_3 <: T }
      { \Gamma \vdash \exsugar{t_1\,\Match\,S_2;\,x,y\,\Rightarrow\,S_3} <: T }
  \end{minipage}

  \begin{minipage}{0.96\linewidth}
    \vs\infrule[SubPi]
      { \Gamma \vdash S_2 <: S_1 \quad \Gamma, x : S_2 \vdash T_1 <: T_2 }
      { \Gamma \vdash \tabs{x}{S_1}{T_1} <: \tabs{x}{S_2}{T_2} }
  \end{minipage}

  \begin{minipage}{0.96\linewidth}
    \vs\infrule[SubNorm]
      { \Gamma \vdash \tnorm{T_1}{T_1'} \quad \Gamma \vdash \tnorm{T_2}{T_2'} \quad
        \Gamma \vdash \untangl{T_1'} <: \untangl{T_2'} }
      { \Gamma \vdash T_1 <: T_2 }
  \end{minipage}

  \end{framed}
  \caption{#1}
  \label{subtypingRulesNew}
  \end{figure}
}

\newcommand\typeNormalizationRules[1]{
  \begin{figure}[!tb]
  \begin{framed}

  \begin{minipage}{0.35\linewidth}
    \infrule[NBase]
      {T \in \{\Any, \exsugar{\List}\}}
      { \Gamma \vdash \tnorm{T}{T} }
  \end{minipage}
  \begin{minipage}{0.55\linewidth}
    \infrule[NSing]
      { \Gamma \vdash \reducesBD{t}{t'} \quad \Gamma \vdash t' \Uparrow \singletonOf{t''}{V} }
      { \Gamma \vdash \tnorm{\singletonOf{t}{U}}{\singletonOf{t''}{V}} }
  \end{minipage}

  \begin{minipage}{0.90\linewidth}
    \vs\infrule[NPi]
      { \Gamma \vdash \tnorm{S}{S'} \quad \Gamma, x\!:\!S' \vdash \tnorm{T}{T'}  }
      { \Gamma \vdash \tnorm{\tabs{x}{S}{T}}{\tabs{x}{S'}{T'}} }
  \end{minipage}

  \begin{minipage}{0.9\linewidth}
    \vs\infrule[NExists1]
      { \Gamma \vdash \tnorm{S}{S'}  \quad  \Gamma, x: S' \vdash \tnorm{T}{T'}  \quad  x \not\in \fv{T} }
      { \Gamma \vdash \tnorm{\texists{x}{S}{T}}{T'} }
  \end{minipage}

  \begin{minipage}{0.9\linewidth}
    \vs\infrule[NExists2]
      { \Gamma \vdash \tnorm{S}{S'}  \quad  \Gamma, x: S' \vdash \tnorm{T}{T'}  \quad  x \in \fv{T} }
      { \Gamma \vdash \tnorm{\texists{x}{S}{T}}{\texists{x}{S'}{T'}} }
  \end{minipage}

  \begin{minipage}{0.90\linewidth}
    \vs\infrule[NCons]
      { \Gamma \vdash \tnorm{T_1}{T_1'} \quad \Gamma \vdash \tnorm{T_2}{T_2'} }
      { \Gamma \vdash \tnorm{\exsugar{\Cons~T_1~T_2}}{\exsugar{\Cons~T_1'~T_2'}} }
  \end{minipage}

  \begin{minipage}{0.90\linewidth}
    \vs\infrule[NMatch1]
      { \Gamma \vdash \reducesBD{\exsugar{t}}{\exsugar{\nil}} \quad \Gamma \vdash \tnorm{T_2}{T_2'} }
      { \Gamma \vdash \tnorm{\exsugar{t\,\Match\,T_2;\,x,y\,\Rightarrow\,T_3}}{T_2'} }
  \end{minipage}

  \begin{minipage}{0.9\linewidth}
    \vs\infrule[NMatch2]
      { \Gamma \vdash \reducesBD{\exsugar{t}}{\exsugar{\cons~t_1~t_2}} \quad
        \Gamma, x\!:\!\singletonOf{t_1}{\Any}, y\!:\!\singletonOf{t_2}{\exsugar{\List}} \vdash \tnorm{T_3}{T_3'} }
      { \Gamma \vdash \tnorm{\exsugar{t\,\Match\,T_2;\,x,y\,\Rightarrow\,T_3}}{T_3'} }
  \end{minipage}

  \begin{minipage}{0.9\linewidth}
    \vs\infrule[NMatch3]
      { \Gamma \vdash \reducesBD{\exsugar{t}}{\exsugar{t'}} \quad
        \text{if neither of the above rules apply} }
      { \Gamma \vdash \tnorm{\exsugar{t\,\Match\,T_2;\,x,y\,\Rightarrow\,T_3}}{\exsugar{t'\,\Match\,T_2;\,x,y\,\Rightarrow\,T_3}} }
  \end{minipage}

  \end{framed}
  \caption{#1}
  \label{typeNormalizationRules}
  \end{figure}
}

\newcommand\termEvaluationRules[1]{
  \begin{figure}[!htb]
  \begin{framed}

  \begin{minipage}{0.96\linewidth}
  \textbf{Evaluation contexts:}\\
  $$
  \begin{aligned}
    \quad \mathcal{E} \quad\defeq{}\quad &
      []~|~%
      \mathcal{E}~t~|~%
      v~\mathcal{E}~|~%
      \cons~\mathcal{E}~t~|~%
      \cons~v~\mathcal{E}~|~%
      \listmatch{\mathcal{E}}{t}{t}~|~%
  \end{aligned}
  $$
  \end{minipage}
  \bigskip

  \begin{minipage}{0.96\linewidth}
  \textbf{Term evaluation:}\\
  \end{minipage}

  \begin{minipage}{0.30\linewidth}
    \infrule[BCtx]
      { \reducesOne{t}{t'} }
      { \reducesOne{\mathcal{E}[t]}{\mathcal{E}[t']} }
  \end{minipage}
  \begin{minipage}{0.40\linewidth}
    \infrule[BApp]
      {~}
      { \reducesOne{(\abs{x}{A}{t})~v }{ t \subst{x}{v}} }
  \end{minipage}

  \begin{minipage}{0.60\linewidth}
    \infrule[BMatchNil]
      {~}
      { \reducesOne{(\listmatch{\nil}{t_1}{t_2})}{t_1} }
  \end{minipage}

  \begin{minipage}{0.90\linewidth}
    \infrule[BMatchCons]
      {~}
      { \reducesOne{(\listmatch{(\cons~v_1~v^\List_2)}{t_1}{t_2})}{t_2 \subst{x}{v_1} \subst{y}{v^\List_2}} }
  \end{minipage}

  \begin{minipage}{0.80\linewidth}
    \vs\infrule[BFixRec]
      { n = n' + 1 }
      { \reducesOne{\fixd{n}{A}{t_1}{t_2}}{t_1 \subst{x}{\fixd{n'}{A}{t_1}{t_2}}} }
  \end{minipage}

  \begin{minipage}{0.60\linewidth}
    \vs\infrule[BFixDefault]
      { n = 0 }
      { \reducesOne{\fixd{n}{A}{t_1}{t_2}}{t_2} }
  \end{minipage}

  \begin{minipage}{0.40\linewidth}
    \vs\infrule[BChooseTop]
      { \fv{v} = \emptyset }
      { \reducesOne{\choos[\Any]}{v} }
  \end{minipage}
  \begin{minipage}{0.45\linewidth}
    \vs\infrule[BChooseList]
      { \fv{v^\List} = \emptyset }
      { \reducesOne{\choos[\List]}{v^\List} }
  \end{minipage}

  \end{framed}
  \caption{#1}
  \label{termEvaluationRules}
  \end{figure}
}

\newcommand\betaDeltaReduction[1]{
  \begin{figure}[tb]
  \begin{framed}

  \begin{minipage}{0.35\linewidth}
    \infrule[BDCtx]
      { \Gamma \vdash \reducesBDOne{t}{t'} }
      { \Gamma \vdash \reducesBDOne{\mathcal{E}[t]}{\mathcal{E}[t']} }
  \end{minipage}
  \begin{minipage}{0.31\linewidth}
    \infrule[BDDelta]
      { \Gamma(x) = \singletonOf{t}{U} }
      { \Gamma \vdash \reducesBDOne{x}{t} }
  \end{minipage}
  \begin{minipage}{0.30\linewidth}
    \infrule[BDBeta]
      { \reducesOne{t}{t'} }
      { \Gamma \vdash \reducesBDOne{t}{t'} }
  \end{minipage}

  \end{framed}
  \caption{#1}
  \label{betaDeltaReduction}
  \end{figure}
}

\newcommand\termsAndTypes[1]{
  \begin{figure}[!htb]
  \begin{framed}
  \begin{minipage}{0.95\linewidth}
  \textbf{Terms and Types of \oursystem:}\\
  $$
  \begin{aligned}
    p, t \quad\defeq{}\quad        &  x~|~\abs{x}{T}{t}~|~t~t~|~\nil~|~\cons~t~t~|~\listmatch{t}{t}{t}~|\\
                            & ~\fixd{n}{T}{t}{t}~|~\choos[B]\\
    S, T, U, V \quad\defeq{}\quad  &  B~|~\singletonOf{t}{T}~|~\tabs{x}{T}{T}\\
    B \quad\defeq{}\quad           &  \Any~|~\List\\
  \end{aligned}
  $$
  \end{minipage}
  \vspace{1.5em}

  \begin{minipage}{0.95\linewidth}
  \textbf{Values:}\\
  \vspace{-2.1em}
  $$
  \begin{aligned}
    v, v^\Any \quad\defeq{}\quad   &  x~|~\abs{x}{T}{t}~|~v^\List\\
    v^\List \quad\defeq{}\quad     &  \nil~|~\cons~v~v^\List
  \end{aligned}
  $$
  \end{minipage}
  \end{framed}
  \caption{#1}
  \label{termsAndTypes}
  \end{figure}
  \vspace{1em}
}

\newcommand\highlight[1]{\colorbox{light-gray}{$\displaystyle {#1}$}}
\newcommand\termsAndTypesLow[1]{
  \begin{figure}[!t]
  \begin{framed}
  \begin{minipage}{0.95\linewidth}
  \textbf{Terms and Types of \oursystemlow:}\\
  $$
  \begin{aligned}
    p, t \quad\defeq{}\quad        &  x~|~\abs{x}{T}{t}~|~t~t~|~\nil~|~\cons~t~t~|~\listmatch{t}{t}{t}~|\\
                            & ~\fixd{n}{T}{t}{t}~|~%
                                \highlight{~\traildot{t}{1}~|~\traildot{t}{2}~|~\traildot{t}{3}}\\
    S, T, U, V \quad\defeq{}\quad  &  B~|~\singletonOf{t}{T}~|~\tabs{x}{T}{T}~|\\
                            & ~\highlight{\Cons~T_1~T_2~|~\tlistmatch{t}{T}{T}~|~\texists{x}{T}{T}~|~\Trail}\\
  \end{aligned}
  $$
  \end{minipage}
  \vspace{1.5em}

  \begin{minipage}{0.95\linewidth}
  \textbf{Values:}\\
  \vspace{-2.3em}
  $$
  \begin{aligned}
    v, v^\Any \quad\defeq{}\quad   &  x~|~\abs{x}{T}{t}~|~v^\List~|~%
                                \highlight{\traildot{v}{1}~|~\traildot{v}{2}~|~\traildot{v}{3}} \\
  \end{aligned}
  $$
  \end{minipage}

  \end{framed}
  \caption{#1}
  \label{termsAndTypesLow}
  \end{figure}
}

\newcommand\chooseEncodingRules[1]{
  \begin{figure}[!t]
  \begin{framed}

  $$
  \begin{aligned}
    \sugarC{}{T} :\ & \text{Type} \rightarrow \text{Type}\\
    \sugarC{}{B} \defeq{}\ & B\\
    \sugarC{}{\singletonOf{t}{T}} \defeq{}\ & \texists{z}{\Trail}{\singletonOf{\sugarC{z}{t}}{\sugarC{}{T}}}
      \hspace{0.5em} \text{ where $z$ is fresh}\\
    \sugarC{}{\tabs{x}{S}{T}} \defeq{}\ & \tabs{z}{\Trail}{\tabs{x}{\sugarC{}{S}}{\sugarC{}{T}}}
      \hspace{0.5em} \text{ where $z$ is fresh}\\
    \sugarC{}{\Cons~T_1~T_2} \defeq{}\ & \Cons~\sugarC{}{T_1}~\sugarC{}{T_2}\\
    \sugarC{}{\tlistmatch{t}{T_2}{T_3}} \defeq{}\ & \tlistmatch{\sugarC{z}{t}}{\sugarC{}{T_2}}{\sugarC{}{T_3}}
      \hspace{0.5em} \text{ where $z$ is fresh}\\
    \\
    \sugarC{p}{t} :\ & \text{Term} \rightarrow \text{Term} \rightarrow \text{Term}\\
    \sugarC{p}{\choos[B]} \defeq{}\ & \textit{\unpack}_{B}~p\\
    \sugarC{p}{\abs{x}{T}{t}} \defeq{}\ & \abs{z}{\Trail}{\abs{x}{\sugarC{}{T}}{\sugarC{z}{t}}}
      \hspace{0.5em} \text{ where $z$ is fresh}\\
    \sugarC{p}{t_1~t_2} \defeq{}\ & t_1'~\traildot{p}{3}~t_2'
      \hspace{0.5em} \text{ where }
        t_1' = \sugarC{\traildot{p}{1}}{t_1} \text{ and }%
        t_2' = \sugarC{\traildot{p}{2}}{t_2}\\
    \sugarC{p}{x} \defeq{}\ & x\\
    \sugarC{p}{\nil} \defeq{}\ & \nil\\
    \sugarC{p}{\cons~t_1~t_2} \defeq{}\ & \cons~\sugarC{\traildot{p}{1}}{t_1}~\sugarC{\traildot{p}{2}}{t_2}\\
    \sugarC{p}{\listmatch{t_1}{t_2}{t_3}} \defeq{}\ &
      \listmatch{\sugarC{\traildot{p}{1}}{t_1}}{\sugarC{\traildot{p}{2}}{t_2}}{\sugarC{\traildot{p}{3}}{t_3}}\\
    \sugarC{p}{\fixd{n}{T}{t_1}{t_2}} \defeq{}\ & \fixd{n}{\sugarC{}{T}\!}{\sugarC{\traildot{p}{1}}{t_1}}{\sugarC{\traildot{p}{2}}{t_2}}\\
  \end{aligned}
  $$

  \end{framed}
  \caption{#1}
  \label{chooseEncodingRules}
  \end{figure}
}

\newcommand\translationToSystemFR[1]{
  \begin{figure}[!t]
  \begin{framed}

  \begin{minipage}{0.98\linewidth}
  \textbf{Translation of terms to System FR:}
  $$
  \begin{aligned}
    \sugar{x} \defeq{}\ & x\\
    \sugar{\abs{x}{T}{t}} \defeq{}\ & \eabs{x}{\sugar{t}}\\
    \sugar{t_1~t_2} \defeq{}\ & \sugar{t_1}~\sugar{t_2}\\
    \sugar{\nil} \defeq{}\ &
      \text{left}~(\Unit + (\Any, \sugar{\List}))(())\\
    \sugar{\cons~t_1~t_2} \defeq{}\ &
      \text{right}~(\Unit + (\Any, \sugar{\List}))((\sugar{t_1}, \sugar{t_2}))\\
    \sugar{\listmatch{t_1}{t_2}{t_3}} \defeq{}\ &
      \text{either{\textscale{.5}{\textunderscore}}match}(\sugar{t_1},
        z \Rightarrow \sugar{t_2},
        z \Rightarrow \sugar{t_3} \subst{x}{\pi_1\,z} \subst{y}{\pi_2\,z})\\
    \sugar{\fixd{n}{X}{t_1}{t_2}} \defeq{}\ &
      \fix(x \Rightarrow \abs{y}{\Nat}{~}
          \text{match}(y,~%
          \sugar{t_2},~%
          y' \Rightarrow \sugar{t_1} \subst{x}{x\,y'})
        )~n\\
  \end{aligned}
  $$
  \end{minipage}
  \vspace{1em}

  \begin{minipage}{0.98\linewidth}
  \textbf{Translation of types to System FR:}
  $$
  \begin{aligned}
    \sugar{\Any} \defeq{}\ & \Any\\
    \sugar{\List} \defeq{}\ & \forall n.~\text{Rec}(n)(X \Rightarrow \Unit + (\Any,X))\\
    \sugar{\singletonOf{t}{T}} \defeq{}\ & \refinementOf{\sugar{T}}{v}{\sugar{t}}\\
    \sugar{\tabs{x}{S}{T}} \defeq{}\ & \tabs{x}{\sugar{S}}{\sugar{T}}\\
    \sugar{\Cons~T_1~T_2} \defeq{}\ & \existsTerm{x_1}{\sugar{T_1}}~\existsTerm{x_2}{\sugar{T_2}}~\singletonOf{\sugar{\cons~x_1~x_2}}{\List}\\
    \sugar{\tlistmatch{t}{T_2}{T_3}} \defeq{}\ &
      \refinementOf{\sugar{T_2}}{t}{\text{left}~()}~\cup\\
      & \existsTerm{y_1}{\Any}~\existsTerm{y_2}{\sugar{\List}}\\
      & \quad
        \refinementOf{\sugar{T_3} \subst{x_1}{y_1} \subst{x_2}{y_2}}{t}{\text{right}~(y_1, y_2)}\\
    \sugar{\texists{x}{S}{T}} \defeq{}\ & \texists{x}{\sugar{S}}{\sugar{T}}
  \end{aligned}
  $$
  \end{minipage}

  \end{framed}
  \caption{#1}
  \label{translationToSystemFR}
  \end{figure}
}

\newcommand\untangleDefinition[1]{
  \begin{figure}[!tb]
  \begin{framed}
  \begin{minipage}{1.0\linewidth}
  \newcommand{\gor}{\,\vee\,}
  \newcommand{\gand}{\,\wedge\,}
$$
\begin{aligned}
\mathcal{U} &: \text{Type} \rightarrow \text{Type}
  \hspace{8.0em}\textbf{\textit{``Untangle all Trail existentials''}}\\
\untangl{\texists{x}{\Trail}{T}}    &\defeq{} \untanglWrap{\text{ps}}{\untangl{T}}
  \quad\hspace{0.6em} \text{where } \text{ps} = \trailsOf{x}{\untangl{T}}\\
\untangl{\texists{x}{S}{T}}         &\defeq{} \texists{x}{\untangl{S}}{\untangl{T}}
  \quad                \text{if } S \not= \Trail\\
\untangl{\tabs{x}{S}{T}}            &\defeq{} \tabs{x}{\untangl{S}}{\untangl{T}}\\
\untangl{\singletonOf{t}{U}}        &\defeq{} \singletonOf{t}{\untangl{U}}
  \hspace{6.02em}
  \untangl{\Cons~T_1~T_2}           \defeq{} \Cons~\untangl{T_1}~\untangl{T_2}\\
\untangl{B}                         &\defeq{} B
  \hspace{3.7em}
  \untangl{\tlistmatch{t}{T_2}{T_3}}  \defeq{} \tlistmatch{t}{\untangl{T_2}}{\untangl{T_3}}\\
\end{aligned}
$$
\vspace{1em}
$$
\begin{aligned}
\mathcal{W}_{\!x} &: \text{Id} \rightarrow 2^\text{Term} \rightarrow \text{Type} \rightarrow \text{Type}
  \hspace{0.5em}\textbf{\textit{``Untangle one Trail existential''}}\\
\untanglWrap{\{\}}{T}               &\defeq{} T\\
\untanglWrap{\{p\} \uplus \text{ps}}{T}  &\defeq{}
  \begin{cases}
    \untanglWrap{\text{ps}}{\texists{y}{B}{T'}}      & \text{if } p \not\in \trailsOf{x}{T'}\\
    \untanglWrap{\text{ps}}{\texists{y}{\Trail}{T''}} & \text{otherwise}
  \end{cases}\\
  where &\ \text{$T'$ is $T$ with all occurrences of $(\unpack_B\,x..p)$ replaced by $y$,}\\
        &\ \text{$T''$ is $T$ with all occurrences of $x..p$ replaced by $y$, and $y$ is fresh.}\\
\end{aligned}
$$
\vspace{1em}
$$
\begin{aligned}
\hspace{2em}
\trailsOf{x}{T} &\subseteq \text{Term}
  \hspace{10.5em}\textbf{\textit{``Collect all trails rooted in x''}}\\
\trailsOf{x}{T} & \text{ is a set of maximal selections $p$ where $x..p$ appears in $T$}\\
\end{aligned}
$$
  \end{minipage}
  \end{framed}
  \caption{#1}
  \label{untangleDefinition}
  \end{figure}
}

%% file: abstract.md
Type-level programming is an increasingly popular way to obtain additional type safety.
Unfortunately, it remains a second-class citizen in the majority of industrially-used programming languages.
We propose a new dependently-typed system with subtyping and singleton types whose goal is to enable type-level programming in an accessible style.
At the heart of our system lies a non-deterministic choice operator.
We argue that embracing non-determinism is crucial for bringing dependent types to a broader audience of programmers, since real-world programs will inevitably interact with imprecisely-typed, or even impure code.
Furthermore, we show that singleton types combined with the choice operator can serve as a replacement for many type functions of interest in practice.
We establish the soundness of our approach using the Coq proof assistant. Our soundness approach models non-determinism using additional function arguments to represent choices. We represent type-level computation using singleton types and existential types that quantify over choice arguments.
To demonstrate the practicality of our type system, we present an implementation as a modification of the Scala compiler.
We provide a case study in which we develop a strongly-typed wrapper for Spark datasets.

%% file: content.tex
\section{Introduction}\label{introduction}

Dependent types have been met with considerable interest from the
research community in recent years. Their primary application so far has
been in proof assistants such as Agda \autocite{agdaNorell2007towards},
Coq \autocite{coqBook} and Idris \autocite{brady2013idris}, where they
provide a sound and expressive foundation for theorem proving. However,
dependent types are still largely absent from general-purpose
programming languages, despite a long history of lightweight approaches
\autocite{xi1998eliminating}. In the context of Haskell, much research
has gone into extending the language to support computations on types,
for instance in the form of functional dependencies
\autocite{functionalDependencies}, type families
\autocite{kiselyov2010fun} and promoted datatypes
\autocite{promotedDatatypes}. These techniques have seen adoption by
Haskell programmers, showing that there is a real demand for such
mechanisms. Furthermore, recent research has explored how dependent
types could be added to the language for the same purpose
\cite{dependentHaskell, dependentHaskellSpec}. In a largely orthogonal
direction, inference for dependent refinement types is reaching
significant maturity
\cite{vazou2018gradual, vazou2017refinement,vazou15bounded}.

Dependently-typed languages often rely on a unified syntax to describe
both terms and types. The simplicity of this approach is unfortunately
at odds with the design of most programming languages, where types and
terms are expressed using separate syntactic categories. Singleton types
provide a simple solution to this problem by allowing every term to be
represented as a type. The singleton type of a term therefore gives us
the most precise specification for that term.

In this paper, we report on our attempt at combining an industrial
mixed-paradigm language, Scala, with dependent types. We offer both a
formalization of our type system and a discussion of the challenges
faced in a practical implementation. It is our hope that the present
paper will serve as a guide for other language implementors interested
in pushing the limit of their type systems by adding dependent types.

Unlike proof assistants, we do not aim to use types as a general-purpose
logic, which would favor designs ensuring totality of functions through
termination checks. Instead, our focus is on improving type safety by
increasing the expressive power of the type system.

We present \oursystem, a dependently-typed calculus with subtyping and
singleton types. The main novelty of our calculus is a new approach to
expressing type-level computation that, at first, seems diametrically
opposed to the purity other systems favor. A new term is added for
non-deterministic choice from a base type, similar to Floyd's choice
operator \autocite{floyd1967nondeterministic}. Designing a sound system
in the presence of non-determinism is challenging. Our solution provides
systematic translation of non-determinism using additional parameters
that are existentially quantified at a syntactically well-defined point.
Consequently, a term in \oursystem may reduce to different values. Our
system generalizes the traditional notion of singleton type: when the
lifted term \(t\) contains a non-deterministic choice, the resulting
type \(\singleton{t}\) denotes the set of values that \(t\) could
possibly reduce to. As a result, our type system is capable of type
computations by manipulating types which are based on terms, but can
nonetheless contain more than a single value. In combination with
subtyping, this allows us to seamlessly integrate with impure, or
imprecisely-typed programs.

Our contributions are as follows:

\begin{itemize}
\item
  We present our calculus \oursystem, which illustrates the novel
  elements of our extension to Scala (\Cref{sec:new-formalism}). The
  type system of \oursystem combines dependent types, subtyping and a
  generalization of singleton types to non-deterministic terms. We
  demonstrate how the interplay of these features allows us to leverage
  term-level programs for type-level computation.
\item
  We provide a soundness proof of \oursystem by reusing reducibility
  semantics of \FR \cite{systemfr2019}. Using its semantics we prove the
  soundness of our rules (\Cref{sec:soundness}). These proofs are
  mechanized using the Coq proof assistant \autocite{coqBook}. The
  formalization is available in the additional materials.
\item
  We show a concrete use-case of our system by implementing it as an
  extension of Scala (\Cref{sec:extending-scala}), and using it to
  develop a strongly-typed wrapper for Apache Spark \autocite{spark}
  (\Cref{sec:use-case}). Thanks to dependent types, we can statically
  ensure the type safety of database operations such as join and filter.
  We compare our implementation with an equivalent implicit-based one
  and show remarkable compilation time savings.
\end{itemize}

\section{Motivating Examples}\label{sec:motivating-example}

We begin by motivating why dependent types are desirable in general
purpose programming, and how one might use them to improve type safety.
In our first example, we design an API that keeps track of database
tables' schemas in the type. We demonstrate how dependently-typed list
operations can be used to compute schemas resulting from join operations
at the type level. Our second example shows how to build a safer version
of the zip operation on lists that only accepts equally-sized arguments.
The examples in this section are written in our dependently-typed
extension of Scala described in \Cref{sec:extending-scala}.

\subsection{Safe Join}\label{safe-join}

As a first step, we show how our system supports type-level programming
in the style of term-level programs. Consider the following definition
of the list datatype, which is standard Scala up the
\lstinline!dependent! keyword:

\begin{lstlisting}
sealed trait Lst { (*\text{\dots}*) }
dependent case class Cons(head: Any, tail: Lst) extends Lst
dependent case class Nil() extends Lst
\end{lstlisting}

\noindent
We can define list concatenation in the usual functional style of
Scala\footnote{Our examples use the indentation-based syntax introduced in Scala 3.0.},
that is, using pattern matching and recursion:

\begin{lstlisting}
sealed trait Lst:
  dependent def concat(that: Lst) <: Lst =
    this match
      case Cons(x, xs) => Cons(x, concat(xs, that))
      case Nil() => that
\end{lstlisting}

\noindent
By annotating a method as \lstinline!dependent!, the user instructs our
system that the result type of \lstinline!concat! should be as precise
as its implementation. Effectively, this means that the body of
\lstinline!concat! is lifted to the type level, and will be partially
evaluated at every call site to compute a precise result type which
\emph{depends} on the given inputs. For recursive \lstinline!dependent!
methods such as \lstinline!concat!, we infer types that include calls to
\lstinline!concat! itself. The \lstinline!<:! annotation lets us provide
an upper bound on \lstinline!concat!'s result type, which will be used
while type checking the method's definition. Finally, by qualifying the
definition of \lstinline!Cons! and \lstinline!Nil! as
\lstinline!dependent! we also allow their constructors and extractors to
be lifted to the type level. Using these definitions, we can now request
the precise type whenever we manipulate lists by annotating the new
\lstinline!val! binding with \lstinline!dependent!:

\begin{lstlisting}
dependent val l1 = Cons("A", Nil())
dependent val l2 = Cons("B", Nil())
dependent val l3 = l1.concat(l2)
l3.size: { 2 }
l3: { Cons("A", Cons("B", Nil())) }
\end{lstlisting}

Enclosing a pure term in braces (\lstinline!{!~\ldots{}~\lstinline!}!)
denotes the singleton type of that term. In the last two lines of this
example we are therefore asking our system to prove that \lstinline!l3!
has size 2 and is equivalent to \lstinline!Cons("A", Cons("B", Nil()))!.

In Scala we often deal with impure or imprecisely-typed code, however.
To integrate with such terms, we provide the \lstinline!choose[T]!
construct. Operationally, we interpret \lstinline!choose[T]! as a
non-deterministic choice from \lstinline!T!, which can be modeled
faithfully on the type level as an existentially quantified inhabitant
of \lstinline!T! in a singleton type. Thus, we equate
\lstinline!{ choose[T] }! to \lstinline!T!, and when typing an impure
term such as \lstinline!Cons(readString(), Nil())! we can assign the
precise type \lstinline!{ Cons(choose[String], Nil()) }!. Returning to
the previous example, this means that even in the presence of impurity,
we can perform useful type-level computation and checking:

\begin{lstlisting}
dependent val l2 = Cons(readString(), Nil())
dependent val l3 = l1.concat(l2)
l3: { Cons("A", Cons(choose[String], Nil())) }
\end{lstlisting}

In a style similar to \lstinline!concat!, we can define remove on
\lstinline!Lst!:

\begin{lstlisting}
sealed trait Lst:
  dependent def remove(e: String) <: Lst =
    this match
      case Cons(head, tail) =>
        if (e == head) tail
        else Cons(head, tail.remove(e))
      case _ => throw new Error("element not found")
\end{lstlisting}

\noindent
Removing \lstinline!"B"! yields the expected result, while trying to
remove \lstinline!"C"! from \lstinline!l3! leads to a \emph{compilation
error}, since the given program will provably fail at runtime.

\begin{lstlisting}
l3.remove("B"): { Cons("A", Nil()) }
l3.remove("C") // Error: element not found
\end{lstlisting}

The lists we defined so far can be used to implement a type-safe
interface for database tables.

\begin{lstlisting}
dependent case class Table(schema: Lst, data: spark.DataFrame):
  dependent def join(right: Table, col: String) <: Table =
    val s1 = this.schema.remove(col)
    val s2 = right.schema.remove(col)
    val newSchema = Cons(col, s1.concat(s2))
    val newData = this.data.join(right.data, col)
    new Table(newSchema, newData)
\end{lstlisting}

\noindent
In this example, we wrap a weakly-typed implementation of Spark's
\lstinline!DataFrame! in the \lstinline!dependent! class
\lstinline!Table!. The first argument of this class represents the
schema of the table as a precisely-typed list. The second argument is
the underlying \lstinline!DataFrame!. In the implementation of
\lstinline!join!, we execute the join operation on the underlying tables
(\lstinline!newData!) and compute the resulting schema corresponding to
that join (\lstinline!newSchema!). By annotating the \lstinline!join!
method as \lstinline!dependent!, the resulting schema is reflected in
the type:

\begin{lstlisting}
dependent val schema1 = Cons("age", Cons("name", Nil()))
dependent val schema2 = Cons("name", Cons("unit", Nil()))
dependent val table1  = Table(schema1, (*\text{\dots}*))
dependent val table2  = Table(schema2, (*\text{\dots}*))
dependent val joined  = table1.join(table2, "name")
joined: { Table(Cons("name", Cons("age", Cons("unit", Nil()))), choose[DataFrame]) }
\end{lstlisting}

\noindent
Reflecting table schemas in types increases type safety over the
existing weakly-typed interface. For instance, it becomes possible to
raise compile-time errors when a user tries to use non-existent columns.
This is an improvement over the underlying Spark implementation that
would instead fail at runtime.

\subsection{Safe Zip}\label{safe-zip}

Our first example demonstrated how dependent methods allow inference of
precise types. Conversely, we can also use singleton types to constrain
method parameters further. In this example, our goal is to write a safer
wrapper for functions like \lstinline!zip! that should only be
applicable to lists of the same length. To accomplish this, we can
constrain the second parameter of \lstinline!zip! as follows:

\begin{lstlisting}
def safeZip(xs: Lst, ys: { sizedLike(xs) }) = unsafeZip(xs, ys)
\end{lstlisting}

\noindent
Here we would like \lstinline!{ sizedLike(xs) }! to be inhabited by all
lists of equal length as \lstinline!xs!, regardless of their elements'
values. How can this be achieved, given that \lstinline!sizedLike(xs)!
is a term? By exploiting the non-deterministic interpretation of
\lstinline!choose[T]!, we can provide a succinct definition for
\lstinline!sizedLike!:

\begin{lstlisting}
dependent def sizedLike(xs: Lst) <: Lst =
  xs match
    case Nil() => Nil()
    case Cons(x, ys) => Cons(choose[Any], sizedLike(ys))
\end{lstlisting}

\noindent
Consider, for instance, the meaning of \lstinline!{ sizedLike(xs) }! for
\lstinline!xs = Cons(1, Cons(2, Nil()))!. After reduction, we obtain
\lstinline!{ Cons(choose[Any], Cons(choose[Any], Nil())) }!, which is a
type that represents all lists of size 2. Thus \lstinline!safeZip!
requires that every caller prove that \lstinline!xs! and \lstinline!ys!
are of the same length, which ensures that the underlying implementation
in \lstinline!unsafeZip! will never fail or truncate elements from one
of the lists.

Note that, unlike \lstinline!concat! and \lstinline!remove! that can be
used both on the term and the type level, \lstinline!sizedLike! is here
intended to be used as a type function, but not at runtime.

\subsection{Discussion: From Choices to
Existentials}\label{discussion-from-choices-to-existentials}

Note that \lstinline!{ sizedLike(xs) }! cannot be readily expressed
using existential types and singletons alone. The given list
\lstinline!xs! might be of an arbitrary size, so the number of
existentials needed for all the occurrences of \lstinline!choose[Any]!
is abstract at this point. More specifically,
\lstinline!{ sizedLike(xs) }! can be seen as a union of an unknown
number of existential types: \[
  \singleton{\nil} \cup
  \texists{x_1}{\Any}{\singleton{\cons~x_1\,\nil}} \cup
  \texists{x_1}{\Any}{\texists{x_2}{\Any}{\singleton{\cons~x_1\,(\cons~x_2\,\nil)}}} \cup
  \dots
\] An important contribution of our type system is that it allows users
to express such existential quantifications conditional on the program
unfolding. Our calculus (described in \Cref{sec:new-formalism}) achieves
this by encoding all non-deterministic choices using a single
existential per-type annotation. In particular, we represent
\lstinline!{ sizedLike(xs) }! by \[
  \texists{z}{\Trail}{\singleton{\text{sizedLike'}~z~\text{xs}}}
\] where \((\text{sizedLike'~z~\text{xs}})\) is defined by \[
  \listmatch{\text{xs}}{\nil}{\cons~(\unpack\,\traildot{z}{1})~(\text{sizedLike'}~\traildot{\traildot{z}{2}}{3}~y)}
\] Conceptually, \(z\!:\!\Trail\) corresponds to a map of input values
passed to a deterministic version of the program, i.e.,
\(\text{sizedLike'}\). Programs resulting from our encoding are pure and
deterministic, so we can perform equational reasoning and apply
well-understood techniques for designing sound type systems. At the same
time, our encoding is adequate with respect to non-determinism (which,
in turn, can approximate other language features). In our example,
\((\unpack\,\traildot{z}{1})\) extracts the value at index \(.1\) from
the input \(z\). Note that using the argument of the recursive call,
\(\traildot{\traildot{z}{2}}{3}\), we ensure that each invocation of
\lstinline!choose[T]! in the original program is translated with a
different index (\Cref{subsec:choose-encoding}). This is necessary for
\(\text{sizedLike'}\) to faithfully model the original
(non-deterministic) \(\text{sizedLike}\), in the sense that each
invocation of \lstinline!choose[T]! can be mapped to a different value.
For instance, when \(xs\) is a concrete list of two elements, we end up
with a type encoded as \[
  \texists{z}{\Trail}{ \singleton{\cons~(\unpack\,\traildot{z}{1})~(\cons~(\unpack\,\traildot{\traildot{\traildot{z}{2}}{3}}{1})~\nil)}}
\] which, given our interpretation of the \(\Trail\) type, selections
like \(\traildot{z}{1}\), and the \(\unpack\) operation, is equivalent
to all the lists of two elements.

During type checking, we explicitly eliminate the references to
\(\unpack\) and replace them by fresh existentials: \[
  \texists{x_1}{\Any}{\texists{x_2}{\Any}{\singleton{\cons~x_1~(\cons~x_2~\nil)}}}
\] That is, we ``untangle'' individual existentials that had previously
been tied up together (\Cref{subsec:untangle}). As part of our overall
soundness proof (\Cref{sec:soundness}) we show that untangling produces
equivalent types, which allows us to match different occurrences of
types containing non-deterministic choices when they denote the same
sets of values.

Our type system rules are designed to support type checking with such
existential types and subtyping. We find that it achieves an appealing
combination of expressive power and simplicity: the developers can
denote types using functions that generate sets of values, instead of
manipulating syntactic representations of types. Even if our current set
of type-checking rules does not cover as many type equivalences as we
may wish to have, our soundness approach based on reducibility semantics
and \FR \cite{systemfr2019} allows us to modularly introduce and prove
additional rules in the future.

\termsAndTypes{The terms and types for \oursystem.}
\termEvaluationRules{The term evaluation rules and evaluation contexts.}

\section{Our System}\label{sec:new-formalism}

We present a calculus and a type system that capture the core mechanisms
required for type-level computation in a dependently-typed language with
subtyping. While an implementation on top of Scala must operate in the
presence of a much more general subtyping relation, our formalism does
not cover all the features of Scala's type system. In the following
section, we introduce a functional language with primitives for
operating on Lisp-like lists, which gives similar power as the closed
type hierarchies that our Scala-implementation can reason about. An
extension to other algebraic data types should be straightforward. Our
calculus also supports non-deterministic choice from base types and
\(\Any\). This choice operator allows us, on the one hand, to model
imprecisely-typed functions and, on the other hand, to emulate
type-level computation.

\subsection{Syntax and Semantics}\label{subsec:syntax-and-semantics}

The terms and types of our calculus, \oursystem, are defined in
\Cref{termsAndTypes}. We consider terms and types equivalent up to
alpha-renaming. As usual, variables are named \(x, y\) or \(z\). We
denote the set of free variables of a term \(t\) by \(\fv{t}\). Our
language contains first-class functions and constructors for lists,
along with pattern matching, and a fixpoint combinator. Programs in our
language always terminate because our fixpoint combinator is bounded to
a maximum recursion depth and returns a default value otherwise. In
\(\fixd{n}{T}{t_1}{t_2}\), \(n\) corresponds to the maximum recursion
depth, \(t_1\) is the body of \(\fix\) and \(t_2\) is the default value.
We expect that our approach extends to more general solutions, for
example, requiring proofs of termination as in most dependently-typed
languages \cite{agdaNorell2007towards, coqBook}, or controlling
reduction on the type level using iso-types \autocite{yang2016unified}.

The small-step operational semantics given in \Cref{termEvaluationRules}
is mostly standard, save for two aspects. First, term evaluation does
not get stuck on variables (we include them among the values \(v\)) and
behaves non-deterministically on the term \(\choos[B]\), which evaluates
to an arbitrary value of type \(B\) (i.e., base types or \(\Any\)).
Unlike many other dependently-typed systems, this allows us to express
more than just purely-functional programs, as \(\choos[B]\)
conservatively models a term lacking referential transparency. Second,
\(\choos[B]\) allows us to model the situation in which parts of our
program may be pure, but are typed in a less precise manner.

Besides the dependent products usually found in dependently-typed
languages, we also include singleton types
\autocite{hayashi1991singleton}, denoted \(\singletonOf{t}{U}\), which
are inhabited only by terms observationally equivalent to \(t\). The
\emph{underlying} type \(U\) provides an upper bound for the singleton
type and is used to guide type inference. For instance, \(\nil\) can be
typed precisely as \(\singletonOf{ \nil }{\List}\). The identity
function on \(\Any\) can be typed as: \[
  \singletonOf{ \abs{x}{\Any}{x} }{\tabs{x\,}{\,\Any}{\singletonOf{x}{\Any}}}
\] For better legibility we often write singletons without their
underlying types: \(\singleton{\abs{x}{\Any}{x}}\).

When used in a type annotation, \(\choos[B]\) existentially quantifies
over an arbitrary value in \(B\). As a result, the base type \(\List\)
is equivalent to \(\singleton{ \choos[\List] }\) which in turn allows us
to express the type of non-empty lists as follows: \[
  \singleton{ \cons~(\choos[\Any])~(\choos[\List]) }
\] During type checking, our system rewrites \(\choos[B]\) to explicit
existential quantifications, that are not available in the surface
syntax. Internally, we end up with the following type for the above
example: \[
  \texists{x_1}{\Any}{\texists{x_2}{\List}{\singleton{ \cons~x_1~x_2 }}}
\] Semantically, this type corresponds to the infinite union over all
elements \(x_1\!\!\!:\!\Any, x_2\!\!:\!\List\) of
\(\singleton{\cons~x_1~x_2}\). As a first step towards representing the
impure \(\choos[B]\) construct, we translate programs in \oursystem to a
deterministic language, as described below.

\subsection{Lowering to a Deterministic
Language}\label{subsec:choose-encoding}

In this section, we detail how we eliminate the non-deterministic
\(\choos[B]\) construct. The essence of our translation is to collect
all the choices that a non-deterministic execution might need and turn
them into an input argument of a deterministic version of the program.
Our translation is therefore analogous to a translation from a
non-deterministic Turing machine to a deterministic machine that acts as
the corresponding verifier \cite[Theorem 7.20 on p. 294]{sipser2013}.

The encoding is performed before type checking, and as a consequence
\(\choos[B]\) is absent from subsequent typing rules. Depending on the
context where \(\choos[B]\) occurs, it takes on different meanings. In
the context of terms, \(\choos[B]\) refers to a specific value in \(B\),
picked non-deterministically during program execution. When invoked from
inside a singleton type, such as in \(\singleton{\choos[B]}\), our
translation will give it the meaning of all values in \(B\). This result
arises due to existential quantification over choices, which the
translation introduces independently for each type annotation in the
program.

We define a lowering from \oursystem, the surface language, to
\oursystemlow, which we then use in subsequent type checking. In
\Cref{termsAndTypesLow} we give the terms and types of
\oursystemlow with the differences to \oursystem highlighted in gray.
First, note the absence of \(\choos[B]\), which is eliminated by the
lowering. We include types for list constructors (\(\Cons~T_1~T_2\)) and
matches. The type for matches, \(\tlistmatch{t}{T_2}{T_3}\), represents
either \(T_2\) or the substituted form of \(T_3\), depending on the
value of \(t\). These types are used later, during type inference, and
guide subtyping. The other additions, i.e., existential types, the base
type \(\Trail\), and selections on trails, \(\traildot{t}{n}\), are
discussed below in the lowering step.

\termsAndTypesLow{The terms and types in \oursystemlow. Constructs not present in \oursystem are marked in \highlight{\text{gray}}.}

\subsubsection{Encoding
choose{[}T{]}}\label{subsubsec:choose-encoding-properties}

Lowering produces a deterministic program that, thanks to an extra
parameter, captures all of the potential behaviors of the original
(non-deterministic) program. We express the lowered program as a
function of \emph{trails}. Intuitively, a trail \(\tau\) contains all
the information necessary to recover the non-deterministic choices made
in a concrete execution of the original program.

Given a program \(t\) in \oursystem, the lowering yields
\(t_f = \abs{z_\alpha}{\Trail}{\sugarC{z_\alpha}{t}}\) in \oursystemlow,
which encodes the behavior of \(t\) as a pure function. That is, for any
given (potentially non-deterministic) reduction resulting in \(v\),
there exists a trail \(\tau\) such that
\(\reduces{(t_f~\tau)}{\sugarC{}{v}}\)\footnote{We omit the trail argument in $\sugarC{}{v}$, as it is irrelevant when translating values, which can only contain $\choos[B]$ underneath lambdas.}.

In \Cref{chooseEncodingRules} we describe the transformation of terms,
\(\sugarC{p}{t}\), and the transformation of types, \(\sugarC{}{T}\). At
its core, \(\sugarC{p}{t}\) replaces each invocation of \(\choos[B]\) by
an application of a function \(\unpack_B\) to a trail. Given the
original program, one of its non-deterministic executions can be
characterized by a mapping from every invocation of \(\choos[B]\) to the
resulting value in \(B\). With respect to our evaluation relation
\(\rightarrow_\beta\), such a mapping can be obtained by recording the
sequence of non-deterministic choices in \rulename{BChooseTop} and
\rulename{BChooseList}. The initial trail \(z_\alpha\) used to evaluate
the lowered program corresponds to a complete mapping for some
non-deterministic execution. Throughout the lowered program we build up
\emph{selections} on the initial trail using \(\traildot{t}{n}\), which
correspond to subtrails. Calls to \(\unpack_B\) then use the given
subtrail to return a value. In our translation we take care never to
apply \(\unpack_B\) to the same trail twice: Doing so would incorrectly
constrain the outcome of the corresponding invocations to be coupled
together.

\chooseEncodingRules{The rules for lowering programs in \oursystem to \oursystemlow, yielding a deterministic program without the non-deterministic $\choos[B]$ construct.}

In the translation of abstractions, we create a fresh trail parameter
\(z\), which is then used to translate the function's body. This is
essential, as it ensures that in each function invocation we allow for
different non-deterministic choices. Note that it does not seem feasible
to enumerate all the possible invocations of \(\choos[B]\) statically:
For one, the outcome of a \(\choos[B]\) might influence the control flow
of the original program, and, in general, the length of the execution
may be unbounded. To translate an application, we select on the current
trail and pass it as the additional argument. Extending the selection is
crucial to ensure that recursive calls can be distinguished in their
non-deterministic choices. Consequently, we also adapt types that occur
in the annotations of abstractions and fixpoints using \(\sugarC{}{T}\).
In particular, \(\Pi\)-types are rewritten to account for the
newly-introduced \(\Trail\) parameter.

Note that in \(\sugarC{}{T}\) we do not propagate and extend an existing
trail as we do with \(p\) in \(\sugarC{p}{t}\). When translating a
singleton type \(\singletonOf{t}{U}\) we instead wrap the resulting type
in a fresh existential type \(\texists{z}{\Trail}\), which is used in
the translation of \(t\). This is what gives \(\choos[B]\) its dual
meaning at the type level: Rather than referring to one particular
choice, it encompasses all of them.

Our lowering is related to monadic encodings in the style of Wadler
\autocite{wadler1990comprehending}. Our encoding is simpler than a
typical State monad because we only care about the distinctness of
trails, rather than encoding the evaluation order and threading the
resulting state from one subterm to another.

\subsubsection{Trails, More Carefully}\label{trails-more-carefully}

\newcommand{\update}[3]{\text{update}~#1~#2~#3}

We will now give a more concrete definition of what properties a trail,
and the operations that act upon it, must satisfy. We organize the
sequence of values of a trail \(\tau\) as a ternary tree. Leaves of this
tree contain a value, and a tag that encodes the type of the value.
Consider \(t..p\) as notation for \((\dots(t.n_1)\dots).n_k\), i.e.,
applying a series of selections \(p = .n_1\dots.n_k\) where
\(n_1, \dots, n_k \in \{1,2,3\}\) to \(t\). Given a trail \(\tau\) and
selections \(p\), \(\tau..p\) represents the subtree of \(\tau\) when
selecting the \(n_i\)-th child at the \(i\)-th level of \(\tau\). For
trees \(\tau, \tau'\) and a selection \(p\),
\((\update{\tau}{p}{\tau'})\) replaces the subtree selected by \(p\) in
\(\tau\) by \(\tau'\), so that \((\update{\tau}{p}{\tau'})..p = \tau'\).
The \(\unpack_B: (\tabs{x}{\Trail}{B})\) function returns the value at
the root of the given tree, if the type-tag of the value there encodes
\(B\), and \(\nil\) otherwise.

\subsection{The Type System}\label{subsec:type-system}

\inferenceRules{The inference and checking rules.}

We introduce our type system for \oursystemlow which consists of several
inter-dependent relations:

\begin{itemize}
\tightlist
\item
  type inference and checking (\(\Uparrow\) and \(\Downarrow\) in
  \Cref{inferenceRules}),
\item
  subtyping (\(<:\) in \Cref{subtypingRulesNew}), and
\item
  type normalization (\(\rightarrow_N\) in
  \Cref{typeNormalizationRules}).
\end{itemize}

\noindent
To improve legibility of the rules, we omit well-formedness conditions,
and presume that types are well-formed in the given context. For
singleton types, in particular, we maintain the assumption that for any
\(\singletonOf{t}{U}\) we encounter, \(t\) inhabits \(U\). Similarly,
for every list match type \((\tlistmatch{t}{T_2}{T_3})\) we assume that
\(t\) inhabits \(\List\).

\paragraph{Type inference and underlying types}

\Cref{inferenceRules} presents rules that infer the most precise type
for a given term \(t\). In particular, type inference will yield a
singleton type \(\singletonOf{t}{U}\), if \(t\) is well-typed. For each
construct we attach an upper bound as the singleton's underlying type
\(U\). In \textsc{TAbs}, for instance, we ``tag'' the singleton type
inferred for a function with the corresponding \(\Pi\)-type, and in
\textsc{TCons} we attach a special type \(\Cons~T_1~T_2\) only present
during type checking. The underlying type is used to guide checks in
\textsc{TApp} and various subtyping rules.

In \textsc{TApp}, in particular, we expose and match against the
underlying function type of \(t_1\) using the auxiliary
\(\widen{\cdot}\) function. Our goal here is to check that applying
\(t_1\) to \(t_2\) is safe, as usual, while also maintaining a precise
version of the underlying type. Assuming we inferred
\(V = \singletonOf{t_1}{\tabs{x\,}{\,S}{T}}\) for \(t_1\), \(\widen{V}\)
will yield a \(\Pi\)-type equivalent to \(t_1\) for all \(x\) in \(S\),
i.e., \(\tabs{x}{S}{\singletonOf{t_1\,x}{T}}\). We then substitute the
argument term in the result type, yielding
\(\singletonOf{t_1\,t_2}{T \subst{x}{t_2}}\). In \textsc{TApp},
\textsc{TCons} and \textsc{TFix} we also refer to a type-checking
relation (\(t \Downarrow T\)) which is defined as a shorthand (see
\textsc{TCheck}) for inferring the type of \(t\) and checking against
the expected type \(T\) using the subtyping relation.

\subtypingRulesNew{The subtyping rules.}

\paragraph{Subtyping and type normalization}

The subtyping relation is given in \Cref{subtypingRulesNew}. Rules for
reflexivity (\textsc{SubRefl}), \(\Pi\)-types (\textsc{SubPi}), \(\Any\)
and the \(\List\) base type (\textsc{SubTop}, \textsc{SubCons1}) are
standard. The \(\Cons\) type introduced during inference can be subtyped
covariantly (\textsc{SubCons2}); the type \((\tlistmatch{x}{T_2}{T_3})\)
assigned to matches behaves like a union of \(T_2\) and \(T_3\), while
allowing \(T_3\) to retain variables bound in the pattern
(\textsc{SubMatch}). Using \textsc{SubSing} we can approximate a
singleton type \(\singletonOf{t}{T_1}\) occuring on the left-hand side
by its upper bound \(T_1\) and continue subtyping from there
(\textsc{SubSing}).

\typeNormalizationRules{The type normalization rules.}
\betaDeltaReduction{The rules of beta-delta reduction.}

Our system allows for computation on types to take place during
subtyping. Subtyping rule \textsc{SubNorm} bundles two kinds of
normalizing behavior: We first reduce both sides \(T_1\) and \(T_2\)
using type normalization. We then attempt to replace any newly-exposed
occurrences of \(\unpack_B\) by fresh existentials of type \(B\) via the
untangle function \(\untangl{\cdot}\).

The rules for type normalization are detailed in
\Cref{typeNormalizationRules}. We merely distribute over \(\Pi\)-types,
existentials, and \(\Cons\)-types (\textsc{NPi}, \textsc{NExists1},
\textsc{NExists2}, \textsc{NCons}). Since \(S\) is assumed to be
inhabited in existential types \(\texists{x}{S}{T}\), we eliminate such
quantifications whenever the result type \(T\) does not contain \(x\)
free (\textsc{NExists1}).

Singleton types \(\singletonOf{t}{U}\) may be normalized using
\textsc{NSing}, in which we first reduce \(t\) using beta-delta
reduction (\Cref{betaDeltaReduction}). Beta-delta reduction is defined
as a context-aware extension of beta reduction (seen previously in
\Cref{termEvaluationRules}) with a new rule \textsc{BDDelta}, which
allows the elimination of variables whose precise definition is known
from the context (a similar evaluation relation is found in
\autocite{courant2003strong}). Delta reduction steps may lead the
underlying type \(U\) to go out-of-sync with the newly computed term
\(t'\). For instance, given the context
\(\Gamma = x : \singletonOf{\nil}{\List}\) and the type
\(\singletonOf{x}{\Any}\), if we were to normalize using the beta-delta
reduction \(\reducesBD{x}{\nil}\) alone, we would arrive at
\(\singletonOf{\nil}{\Any}\). We can improve upon this --- and in fact
might rely on it in later subtyping queries --- by redoing type
inference on \(t'\), yielding a singleton type with a better bound (in
our example, \(\singletonOf{\nil}{\List}\)).

The rules for match allow reduction of \((\tlistmatch{t}{T_2}{T_3})\)
depending on the beta-delta reduction of \(t\). That is, we normalize to
\(T_2\) when \(t = \nil\) (\textsc{NMatch1}), and to \(T_3\) when \(t\)
is a \(\cons\) (\textsc{NMatch2}). In the latter case, we add
precisely-typed bindings that allow for \(x\) and \(y\) to be
\(\delta\)-reduced during the normalization of \(T_3\). If \(t\) does
not fit either case, we instead normalize to a type that incorporates
the reduced \(t\).

\paragraph{Subtyping existential types}

Existentials only enter the program when lowering type annotations in
\oursystem to \oursystemlow, and in \textsc{SubNorm} via
\(\untangl{\cdot}\). When encountered on the left-hand side, existential
types are eliminated by adding \(x\!:\!S\) to the context
(\textsc{SubExistsLeft}). When an existential occurs on the right-hand
side, we try to guess a valid instantiation \(t\) for \(x\)
(\textsc{SubExistsRight}). The subroutine that guesses \(t\) is modelled
abstractly by \(\text{solve}_x(T_1, S, T_2)\), which is expected to
return a singleton \(\singletonOf{t}{U}\). We make no assumptions on the
implementation of \(\text{solve}_x\), but verify that the outcome is a
valid solution by checking that it conforms to \(S\) and makes the
instantiated right-hand side a super-type of the left-hand side. In
\Cref{subsec:stainlessfit-impl} we discuss one possible concrete
implementation of \(\text{solve}_x\).

\subsection{Untangling Trails}\label{subsec:untangle}

In \Cref{subsec:choose-encoding} we explained how to translate the
non-deterministic \(\choos[B]\) construct into an application of
\(\unpack_B\) to a trail. Therefore, during type checking, we often face
subtyping queries involving applications of \(\unpack_B\) on the
right-hand side. For instance, when checking the program \[
  (\abs{x}{\singleton{\cons~\choos[\Any]~\choos[\List]}}{x})~(\cons~\nil~\nil)
\] we will encounter the following subtyping query: \[
  \singleton{\cons~\nil~\nil} <: \texists{z}{\Trail}{\singleton{\cons~(\unpack_\Any~\traildot{z}{1})~(\unpack_\List~\traildot{z}{2})}}
\] Though the right-hand side is an existential type, this query cannot
be solved by \rulename{SubExistsRight} directly, unless the
\(\text{solve}\) subroutine possesses some deep knowledge about
\(\unpack_B\). That is, a priori it is not evident that there exists a
trail \(z\) such that both \((\unpack_\Any~\traildot{z}{1})\) and
\((\unpack_\List~\traildot{z}{2})\) reduce to \(\nil\).

Using the properties on trails and \(\unpack_B\) introduced in
\Cref{subsubsec:choose-encoding-properties} we can prove that the type
on the right-hand side is, in fact, equivalent to an explicitly
quantified version, i.e., \[
  \texists{z}{\Trail}{\singleton{\cons~(\unpack_\Any~\traildot{z}{1})~(\unpack_\List~\traildot{z}{2})}} ~=~
  \texists{x_1}{\Any}{\texists{x_2}{\List}{\singleton{\cons~x_1~x_2}}}
\] To see why the inclusion from left to right holds, consider any trail
\(\tau\) with values \(v_1^\Any\) and \(v_2^\List\) stored at indices
\(.1\) and \(.2\). We can thus instantiate \(x_1\) and \(x_2\) to
\(v_1^\Any\) and \(v_2^\List\), to obtain the same term on both sides.
From right to left we can construct a tree containing the values of
\(x_1\) and \(x_2\) at indices \(.1\) and \(.2\). This same reasoning
can be applied to all functions of \(\Trail\) that we encounter after
our lowering to \oursystemlow. Furthermore, it generalizes to an
arbitrary number of selections on \(z\), as long as the selections are
not prefixes of one another, which is ensured by our lowering step.

\untangleDefinition{The untangle function $\mathcal{U}$ and additional auxiliary functions.}

To exploit this property, we define the untangle function
\(\untangl{\cdot}\), which transforms the left-hand side of the equality
above to the right-hand side. We use \(\untangl{\cdot}\) during
normalization in \rulename{SubNorm}. In our example, this leads to a
simpler subtyping query: \[
  \singleton{\cons~\nil~\nil} <: \texists{x_1}{\Any}{\texists{x_2}{\List}{\singleton{\cons~x_1~x_2}}}
\] At this point we can apply \rulename{SubExistsRight} twice, which
could find valid assignments for both \(x_1\) and \(x_2\) (i.e.,
\(\nil\)) using straightforward unification.

The definition of \(\untangl{T}\) is given in \Cref{untangleDefinition}.
Given the conditions on trails mentioned above, we prove untangling
always yields equivalent types (see \Cref{sec:soundness}).

\subsection{From Rules to Algorithms}\label{subsec:stainlessfit-impl}

The type system we presented above comes close to being algorithmic. All
of the rules for type inference and most of the rules for subtyping and
type normalization are already syntax-directed. To derive the
normalization of an existential type one has to choose between
\rulename{NExists1} and \rulename{NExists2}, but only one of them will
ever succeed due to the condition on \(x\) being free in \(T'\). It is
therefore straightforward to formulate these cases as a single,
effective rule. In the remainder, we note two more substantial
adjustments that are needed for an effective formulation of our type
system.

The first adjustment is to only apply \rulename{SubNorm} (type
normalization) at the very beginning of subtyping queries (for example,
in \rulename{TCheck}), and before any subderivation that adds a binder
to the context (for example, in the second premise of \rulename{SubPi}).
Applying \rulename{SubNorm} must remain optional, however, since forcing
normalization can lead to subderivations that grow \emph{ad infinitum},
for instance by normalizing under matches and re-entering type inference
in \rulename{NSing}. Beyond making \rulename{SubNorm} optional, in
practice it is useful to allow for a fast path in subtyping. Given a
subtyping query \(T_1 <: T_2\), one can first try to prove a stronger
subtyping relation, where the left-hand side \(T_1\) is approximated by
\(\widen{T_1}\). We found that this greatly reduces the need for complex
subtyping derivations, e.g., when checking against the \(\List\) base
type in \rulename{TCons} and \rulename{TMatch}.

The second adjustment lies in \rulename{SubExistsRight}, where we
require a procedure for \(\text{solve}_x\). In principle,
\(\text{solve}_x(T_1, S, T_2)\) could do arbitrarily deep reasoning
about the involved types, but our experience shows that a
unification-like procedure is sufficient for the use cases presented in
this paper. We experimented with a particularly simple variant:
\(\text{solve}_x\) performs a separate subtyping query which constrains
\(x\) in a greedy manner using a modified version of \rulename{SubRefl}.
In this modified rule, the computation of (syntactic) type equality
looks for any appearances of \(x\). If \(x\) appears on either side of
the comparison, the corresponding term on the other side is picked as a
greedy solution of \(\text{solve}_x\). This naive syntactic approach can
result in an instantiation that is not well-formed in the original
context \(\Gamma\), in which case we simply fail. One could, of course,
try to incrementally improve this approach by trying to rewrite \(t\) to
an equivalent term well-formed in \(\Gamma\). It would be interesting to
explore a more general constraint-solving approach.

\section{Soundness}\label{sec:soundness}

In this section we discuss the formal soundness proof for \oursystemlow.
As a starting point, we use \FR \cite{systemfr2019}, a calculus that was
recently presented as a foundation for the Stainless program verifier
\autocite{stainless}. Our formalization in Coq is available as
additional material in the submission. We give an embedding
\(\sugar{\cdot}\) of \oursystemlow terms and types into \FR in
\Cref{translationToSystemFR}.

\translationToSystemFR{The embedding of \oursystemlow terms and types into \FR.}

\paragraph{Embedding terms}

Functions and applications are represented trivially using \FR's lambda
abstractions and applications, which behave identically to ours. Lists
are encoded in the typical way as a sum of unit for \(\nil\) and a pair
of a head and a tail for \(\cons\).

Our embedding of fix ensures trivial termination, following the bounded
recursion behavior of \oursystem. While we believe that the addition of
general recursion to \oursystem would not present a problem (as
discussed before in \Cref{subsec:syntax-and-semantics}), \FR is
normalizing, and thus prevents us from lifting this restriction in our
current formalization.

\paragraph{Embedding types}

\(\Pi\)-types along with existential types are represented trivially.
The type of lists is expressed in the usual way through a recursive
type. A singleton type \(\singletonOf{t}{T}\) is encoded using the type
\(\refinementOf{T}{v}{t}\) of \FR, which represents all values in \(T\)
that are observationally equivalent to \(t\). Observational equivalence
is supported as a type in the current open-source formalization of \FR,
even though this type was not supported in the paper
\cite{systemfr2019}. The \((\Cons~T_1~T_2)\) type of \oursystemlow is
translated by existentially quantifying over any combination of values
in \(T_1\) and \(T_2\). For the type of matches,
\((\tlistmatch{t}{T_2}{T_3})\), we take the union of each of \(T_2\)'s
and \(T_3\)'s interpretation, conditional on whether the scrutinee \(t\)
reduces to \(\nil\) or a \(\cons\).

\newcommand{\denotOf}[1]{[\![ #1 ]\!]_v}

Given that \FR assigns a reducibility semantics to its types, our
embedding also affords us with denotations for all the types of
\oursystemlow. That is, given the set of reducible values
\(\denotOf{T}\) of type \(T\) in \FR, the meaning of a type \(T'\) in
\oursystemlow is given by \(\denotOf{\sugar{T'}}\). For instance,
\(\denotOf{\sugar{\List}} = \{ \cons~v_1~(\dots(\cons~v_n~\nil) \dots) \mid n \ge 0, \forall i.\, v_i \in \denotOf{\Any} \}\).
Existential types \(\denotOf{\texists{x}{S}{T}}\) are the union of all
\(\denotOf{T \subst{x}{s}}\) for all \(s \in \denotOf{S}\).

\paragraph{Formalized soundness statement}

Using the above embedding, we have proved that all of the rules for type
inference, subtyping and type normalization presented in
\Cref{subsec:type-system} are admissible with respect to the
reducibility semantics of types. We built our mechanization on top of
\FR's existing Coq formalization. The respective lemmas are proven under
the additional assumptions given to us via the well-formedness rules
mentioned in \Cref{subsec:type-system}. Namely, the following are
assumed to hold:

\begin{itemize}
\tightlist
\item
  In rules for type inference, subtyping, and type normalization, we
  require well-formedness in the current context and inhabitedness for
  singleton and list match types.
\item
  During delta-beta reduction, we require terms to be normalizing in the
  current context.
\item
  Trails and their operations are kept abstract and specified using
  axioms in file \lstinline!Trail.v!.
\end{itemize}

\noindent The entirety of our definitions and proofs consists of
\textasciitilde{}\(7\)k lines of Coq in addition to the previous
development of \FR soundness, which consisted of
\textasciitilde{}\(20\)k lines.

We can thus state soundness for \oursystemlow programs in terms of the
reducibility judgment \(\Gamma \vDash t\!:\!T\) of \FR. The latter holds
when, for all substitutions \(\gamma\), such that for all
\((x, S) \in \Gamma\) we have \(\gamma(x) \in \denotOf{\gamma(S)}\),
\(\gamma(t) \in \denotOf{\gamma(T)}\). Let \(\sugar{\Gamma}\) be the
context with all \oursystemlow types embedded into System~FR types.

\begin{thm}[Soundness]
Given a context $\Gamma$ and a \oursystemlow term $t$, if type inference yields a type $T$, then $t$ is reducible at that type.
That is, if $\Gamma \vdash t \Uparrow T$ holds, then $\sugar{\Gamma} \vDash \sugar{t}\!:\!\sugar{T}$.
\end{thm}

Note that the traditional notion of type safety for \(t\) follows, i.e.,
well-typedness of \(t\) implies the existence of value \(v\) such that
\(\reduces{t}{v}\), since \(\sugar{t}\) is normalizing exactly when
\(t\) is. Similarly, using the correspondence of \oursystemlow programs
to non-deterministic \oursystem programs after lowering (see
\Cref{subsec:choose-encoding}), we get type safety for \oursystem.

\section{Implementation}\label{sec:extending-scala}

In this section we give an overview of how we extended Scala with
dependent types. This development was an experiment to explore the
feasibility of adding dependent types in Scala. We implemented our
prototype as an extension of Dotty, the reference compiler for future
versions of the Scala language. Our presentation focuses on several
facets of the implementation that are not reflected in the formalism
presented in \Cref{sec:new-formalism}.

On a syntactic level, our Scala extension consists of three additions:

\begin{itemize}
\tightlist
\item
  the singleton types syntax \lstinline!{ t }!,
\item
  the \lstinline!dependent! modifier for methods, values and classes,
\item
  the \lstinline!choose[T]! construct.
\end{itemize}

\noindent
The newly-introduced singleton type syntax enables a subset of Scala
expressions to be used in types. This subset approximately corresponds
to the core functional subset of Scala, plus the \lstinline!choose[T]!
construct, as illustrated in \oursystem. Within this subset, the main
differences between our formalism and implementation lie in the handling
of pattern matching.

\subsection{Pattern Matching}\label{pattern-matching}

Pattern matching in Scala supports a wide range of matching techniques
\autocite{emir2007matching}. For example, \emph{extractor patterns} rely
on user-defined methods to extract values from objects. As a result,
these custom extractors can contain arbitrary side effects. Our
implementation limits the kind of patterns available in types to the two
simplest forms: decomposition of case classes and the
type-tests/type-casts patterns.

During type normalization, our system evaluates pattern matching
expressions according to Scala's runtime semantics, that is, patterns
are checked top-to-bottom, and type-tests are evaluated using runtime
type information available after type erasure.

For example, consider the following pattern matching expression:

\begin{lstlisting}
s match { case _: T1 => v1 case _: T2 => v2 }
\end{lstlisting}

\noindent
When used in a type, this expression reduces to \lstinline!v1! if the
scrutinee's type is a subtype of \lstinline!T1!. In order to reduce to
\lstinline!v2!, type normalization must make sure \lstinline!T1! and the
scrutinee's type are disjoint, namely that the dynamic type of
\lstinline!s! cannot possibly be smaller than \lstinline!T1!.
Disjointness proofs are built using static knowledge about the class
hierarchy and make use of the guarantees implied by the
\lstinline!sealed! and \lstinline!final! qualifiers, which are Scala's
way of declaring closed-type hierarchies.

\subsection{Two Modes of Type
Inference}\label{subsec:two-type-inference-algorithms}

In order to retain backwards-compatibility, our system supports two
modes of type inference: the precise inference mode which infers
singleton types, and the default inference mode that corresponds to
Scala's current type-inference algorithm. Concretely, users opt into our
new inference mode using the \lstinline!dependent! qualifier on methods,
values, and classes.

When inferring the result type of a \lstinline!dependent! method, our
system lifts the method's body into a type. This lifting will be precise
for the subset of expressions that is representable in types, and
approximative for the rest. When we encounter an unsupported construct,
we compute its type using the default mode, yielding a type
\lstinline!T! which we then integrate in the lifted body as
\lstinline!choose[T]!.

For example, given the following definition:

\begin{lstlisting}
dependent def getName(personalized: Boolean) =
  if (personalized) readString() else "Joe"
\end{lstlisting}

\noindent
our system infers the following result type:

\begin{lstlisting}
{ if (personalized) choose[String] else "Joe" }
\end{lstlisting}

Scala requires recursive methods to have an explicit result type, and
this restriction also applies to \lstinline!dependent! methods. However,
in the case of a \lstinline!dependent! method, an explicit result type
is only used as an upper bound for the actual precise result type and
will only be used to type-check the method's body. At other call sites,
the (precise) inferred result type is used. Bounds of dependent methods
are written using a special syntax (\lstinline!<: T!), which emphasizes
the difference from normal result types (\lstinline!: T!).

\subsection{Approximating Side
Effects}\label{approximating-side-effects}

\paragraph{State}

Scala's type system permits uncontrolled side effects in programs. Given
the absence of an effect system, result types of methods do not convey
any information about the potential use of side effects in the method
body. The situation is analogous for \lstinline!dependent! methods.
Thanks to \lstinline!choose[T]! we can still formulate precise result
types when terms depend on the result of side-effectful operations.
Since we uniformly approximate all side effects, we avoid the situation
where a type refers to a value that may be modified during the program
execution. For instance, if \lstinline!z! is a mutable integer variable,
we will never introduce \lstinline!z! in a singleton type, but we can
still assign a better type than \lstinline!Lst! to an expression like
\lstinline!Cons(z, Nil())!, that is,
\lstinline!{ Cons(choose[Int], Nil()) }!.

\paragraph{Exceptions}

Similarly to how we model other side effects, exceptions are
approximated in types. Our type-inference algorithm uses a new error
type, \lstinline!Error(e)!, which we infer when raising an exception
with \lstinline!throw e!. Exception handlers are typed imprecisely using
the default mode of type-inference. Exceptions thrown in statement
positions are not reflected in singleton types, since the type of
\lstinline!{e1; e2}! is simply \lstinline!{ e2 }!. However, exceptions
thrown in tail positions (such as in remove from
\Cref{sec:motivating-example}) can lead to types normalizing to
\lstinline!Error(e)!. In these cases, our type system can prove that the
program execution will encounter exceptional behavior, and reports a
compilation error. This approach is conservative in that it might reject
programs that recover from exceptions. Also note that this is a sanity
check, rather than a guarantee of no exceptions occurring at runtime.
That is, depending on which rules are used during subtyping, it is
possible to succeed without entering type normalization, resulting in
such errors going undetected. Despite these shortcomings, our treatment
of exceptions results in a practical way to raise compile-time errors.
It would be interesting to explore the addition of an effect system to
our Scala extension and formalization.

\subsection{Virtual Dispatch}\label{virtual-dispatch}

Our extension does not model virtual dispatch explicitly in singleton
types. Instead, the result type of a method call
\lstinline!t.m(!\ldots{}\lstinline!)! is always the result type of
\lstinline!m! in \lstinline!t!'s static type. Consequently,
\lstinline!dependent! methods effectively become \lstinline!final!,
given that only a provably-equivalent implementation could be used to
override it.

Special care must be taken when an imprecisely-typed method is
overridden with a \lstinline!dependent! one. In this situation, the
result type of a method invocation can lose precision depending on type
of the receiver. Calls to the \lstinline!equals! methods are a common
example of this: \lstinline!equals! is defined at the top of Scala's
type hierarchy as referential equality and can be overridden
arbitrarily. Given a class Foo with a \lstinline!dependent! overrides of
\lstinline!equals!, calls to \lstinline!Foo.equals(Any)! and
\lstinline!Any.equals(Foo)! are not equivalent; the former precisely
reflects the equality defined in \lstinline!Foo! whereas the latter
merely returns a \lstinline!Boolean!.

\subsection{Termination}\label{termination}

We distinguish two important aspects of termination.

The first question is whether type-checked programs are guaranteed to
terminate. For simplicity, our work side-steps this question, requiring
bounds for recursion. A more general solution would be to compute or
infer such bounds using measure functions, as done in
\FR \cite{systemfr2019}. Another approach would be to extend our
translation of non-determinism to permit non-termination. We consider
this aspect orthogonal to the objectives of this paper. Our work targets
general-purpose programming language whose type safety is defined with
regards to its runtime semantics and that may include non-terminating
interactive computations.

The second question is termination of our type checker. Non-termination
of type checking implies that the type checker can give three possible
answers, ``type correct'', ``type incorrect'' or ``do not know'' (or
timeout). Treating ``do not know'' as ``type incorrect'' makes the
non-termination unproblematic from a soundness perspective. A similar
argument is made for other dependently-typed languages with unbounded
recursion, such as Dependent Haskell \autocite{dependentHaskell} or
Cayenne \autocite{cayenne}. In practice, our system deals with infinite
loops using a fuel mechanism. Every evaluation step consumes a unit of
fuel, and an error is reported when the compiler runs out of fuel. The
default fuel limit can be increased via a compiler flag to enable
arbitrarily long compilation times.

\section{Use Case}\label{sec:use-case}

In this section, we extend the motivating example presented in
\Cref{sec:motivating-example} by building a type-safe interface for
Spark datasets. We use dependent types to implement a simple
domain-specific type checker for the SQL-like expressions used in Spark.
We then compare the compilation time of our dependently-typed interface
against an equivalent encoding based on implicits.

\subsection{A Type-Safe Database
Interface}\label{a-type-safe-database-interface}

The type-safe interface presented in this section illustrates the
expressive power of our system and is implemented purely as a library.
For brevity, our presentation only covers a small part of Spark's
dataset interface, but the approach can be scaled to cover that
interface in its entirety. The type safety of database queries is a
canonical example and has been studied in many different settings
\cite{leijen1999domain, kazerounian2019type, linq, ur}.

The example built in \Cref{sec:motivating-example} uses lists of column
names to represent schemas. A straightforward improvement is to also
track the type of columns as part of the schema. Instead of using column
names directly, we introduce the following \lstinline!Column! class with
a phantom type parameter \lstinline!T! for the column type, and a field
\lstinline!name! for the column name:

\begin{lstlisting}
dependent case class Column[T](name: String) { (*\text{\dots}*) }
\end{lstlisting}

Table schemas become lists of \lstinline!Column!-s and thereby gain
precision. The definition of \lstinline!join! given in
\Cref{sec:motivating-example} can be adapted to this new schema encoding
to prevent joining two tables that have columns with matching names but
different types.

A large proportion of the weakly-typed Spark interface is dedicated to
building expressions on table columns. Such expressions can currently be
built from strings, in a subset of SQL, or using a Scala DSL which is
essentially untyped.

The lack of type safety for column expressions can be particularly
dangerous when mixing columns of different types. The pitfall is caused
by Spark's inconsistency: depending on types of columns and operations
involved, programs will either crash at runtime, or, more dangerously,
data will be silently converted from one type to another.

By keeping track of column types it becomes possible to enforce the
well-typedness of column expressions. As an example, consider the
following Spark program:

\begin{lstlisting}
table.filter(table.col("a") + table.col("b") === table.col("c"))
\end{lstlisting}

We would like our interface to enforce the following safety properties:

\begin{itemize}
\tightlist
\item
  Columns \(a\), \(b\) and \(c\) are part of the schema of
  \lstinline!table!.
\item
  Addition is well-defined on columns \(a\) and \(b\).
\item
  The result of adding columns \(a\) and \(b\) can be compared with
  column \(c\).
\item
  The overall column expression yields a \lstinline!Boolean!, which
  conforms to filter's argument type.
\end{itemize}

Automatic conversions during equality checks can be prevented by
restricting column equality to expressions of the same type
\lstinline!T!:

\begin{lstlisting}
dependent case class Column[T](k: String):
  def ===(that: Column[T]): Column[Boolean] = Column(s"(${this.k} === ${that.k})")
\end{lstlisting}

Addition in Spark is defined between numeric types and characters. The
result type of an addition depends on the operand types. For numeric
types, Spark will pick the larger of the operand types according to the
following ordering: \lstinline!Double > Long > Int > Byte!. The
situation is quite surprising with characters as any addition involving
a \lstinline!Char! will result in a \lstinline!Double!.

Dependent types can be used to precisely model these conversions. We
define a type function to compute the result type of additions:

\begin{lstlisting}
def addRes(a: Any, b: Any) =
  (a, b) match
    case (_: Char, _: Char | Byte | Int  | Long | Double) => choose[Double]
    case (_: Byte, _: Byte | Int  | Long | Double)        => b
    case (_: Int,  _: Int  | Long | Double)               => b
    case (_: Long, _: Long | Double)                      => b
    case (_: Double, _: Double)                           => choose[Double]
    case (_: Byte | Int | Long | Double, _)               => addRes(b, a)
    case _ => throw new Error("incompatible types in addition")
type AddRes[A, B] = { addRes(choose[A], choose[B]) }
\end{lstlisting}

Also note the use of recursion in the second-to-last case, to avoid
duplicating symmetric cases. The \lstinline!AddRes! type can be used to
define a \lstinline!Column! addition that accurately models Spark's
runtime:

\begin{lstlisting}
dependent case class Column[T] private (k: String):
  dependent def +[U](that: Column[U]) <: Column[_] =
    Column[AddRes[T, U]](s"(${this.k} + ${that.k})")
\end{lstlisting}

Allowing programmers to construct \lstinline!Column!-s from string
literals would defeat the purpose of a type-safe interface. Instead,
programmers should extract columns from a \lstinline!Table!'s schema.
For that purpose, we implement the \lstinline!col! method on
\lstinline!Table! and annotate the \lstinline!Column! constructor as
private.

\begin{lstlisting}
dependent case class Table(schema: Lst, data: spark.DataFrame):
  dependent def col(name: String) <: Column[_] =
    dependent def find(key: String, list: Lst) <: Any =
      list match
        case Cons(head: Column[_], tail) =>
          if (head.k == key) head else find(key, tail)
        case _ => throw new Error("column not found in schema")
    find(name, schema)
  dependent def filter(predicate: Column[Boolean]) <: Table =
    new Table(this.schema, this.data.filter(predicate.k))
\end{lstlisting}

The \lstinline!col! method is implemented using a nested dependent
method to find the column corresponding to the given name. Thanks to the
dependent annotation, the type-checker is able to statically evaluate
calls to \lstinline!col!. Assuming the table's schema contains a column
\lstinline!a! of type \lstinline!Int! and columns \lstinline!b! and
\lstinline!c! of type \lstinline!Long!, the compiler will be able to
infer types as follows:

\begin{lstlisting}
val pred = (*\setlength{\fboxrule}{-1pt}\fcolorbox{light-gray}{light-gray}{ \phantom{.} table.col("a") \phantom{.} }*) + (*\setlength{\fboxrule}{-1pt}\fcolorbox{light-gray}{light-gray}{ \phantom{.} table.col("b") \phantom{.}\phantom{.} }*) === (*\setlength{\fboxrule}{-1pt}\fcolorbox{light-gray}{light-gray}{ \phantom{.} table.col("c") \phantom{.}\phantom{.} }*)
// Infers: (*\setlength{\fboxrule}{-1pt}\fcolorbox{light-gray}{light-gray}{\textit{\{ Column[Int]("a") \}}}*)   (*\setlength{\fboxrule}{-1pt}\fcolorbox{light-gray}{light-gray}{\textit{\{ Column[Long]("b") \}}}*)     (*\setlength{\fboxrule}{-1pt}\fcolorbox{light-gray}{light-gray}{\textit{\{ Column[Long]("c") \}}}*)
\end{lstlisting}

\noindent
Given our definitions of column addition and equality, the overall
\lstinline!pred! expression is typed as \lstinline!Column[Boolean]!. All
the safety properties stated above are therefore enforced by the
dependently-typed interface presented in this section.

\subsection{Comparison to an Existing
Technique}\label{comparison-to-an-existing-technique}

  \begin{figure}
    \centering
    \begin{subfigure}[t]{0.5\textwidth}
        \centering\input{benchmarks/concat-graph.tex}
    \end{subfigure}%
    \hspace{-20pt}
    \begin{subfigure}[t]{0.5\textwidth}
        \centering
        \input{benchmarks/join-graph.tex}
    \end{subfigure}
    \caption{Comparing the compilation times of two implementations of list concatenation and join, log.\ scale.}
    \label{measurements}
  \end{figure}
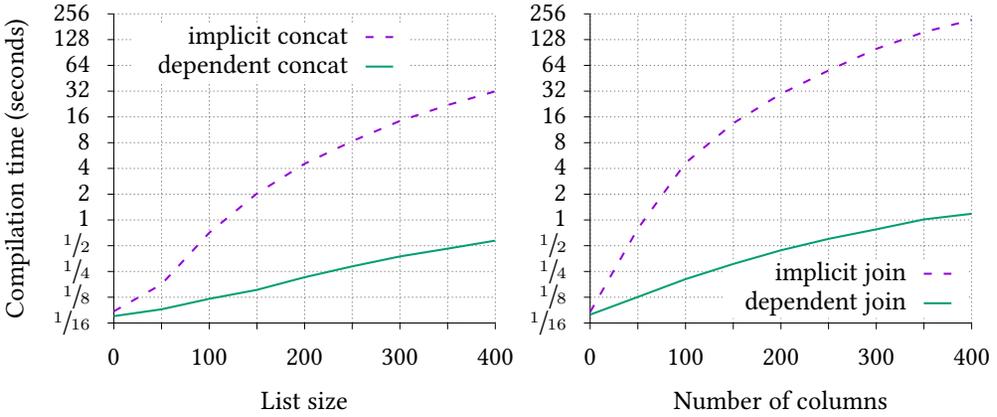

Programmers have managed to find clever encodings that circumvent the
lack of first-class support for type-level programming in many
languages. These encodings can be very cumbersome, as they often entail
poor error reporting and a negative impact on compilation times
\autocite{fakingIt}, \autocite{hlist}. In Scala, implicits are the
primary mechanism by which programmers implement type-level programming
\autocite{implicitPaper}.

Frameless \autocite{frameless} is a Scala library that implements a
type-safe interface for Spark by making heavy use of implicits. Most
type-level computations in this library are performed on the
heterogeneous lists provided by Shapeless \autocite{shapeless}.

We compared the dependently-typed Spark interface presented in this
section against the implicit-based implementation of Frameless. To do
so, we isolated the implicit-based implementation of the
\lstinline!join! operation on table schemas, and compared its
compilation time against the dependently-typed version presented in this
section. To evaluate the scalability of both approaches we generated
test cases with varying schema sizes and compiled each test case in
isolation. A similar comparison is done for list concatenation, which
constitutes a building block of \lstinline!join!.

\Cref{measurements} shows that, in both benchmarks, the
dependently-typed implementation compiles faster than the version with
implicits, and compilation time scales better with the size of the
input. In the join benchmark, we see that the implicit-based
implementation exceeds 30 seconds of compilation time around the 200
columns mark, and continues to grow quadratically. This can be explained
by the nature of implicit resolution, which might backtrack during its
search. The compilation time of the dependently-typed implementation
grows linearly and stays below one second until the 350 columns mark. We
were able to observe similar trends in the concatenation benchmark.
These measurements were obtained by averaging 120 compilations on a warm
compiler, and have been performed on an i7-7700K Processor running
Oracle JVM 1.8.0 on Linux.

\section{Related Work}\label{related-work}

As of today, Haskell is perhaps closest to becoming dependently-typed
among the general-purpose programming languages used in industry.
Haskell's type families \autocite{kiselyov2010fun} provide a direct way
to express type-level computations. Other language extensions such as
functional dependencies \autocite{functionalDependencies} and promoted
datatypes \autocite{promotedDatatypes} are also moving Haskell towards
dependent types. Nevertheless, programming in Haskell remains
significantly different from using full-spectrum dependently-typed
languages. A significant difference is that Haskell imposes a strict
separation between terms and types. As a result, writing
dependently-typed programs in Haskell often involves code duplication
between types and terms. These redundancies can be somewhat avoided
using the singletons package \autocite{eisenberg2012dependently}, which
uses meta-programming to automatically generate types from datatypes and
function definitions.

In the context of Haskell, Eisenberg's work on Dependent Haskell
\autocite{dependentHaskell} is closest to ours, in that it adds
first-class support for dependent types to an established language, in a
backwards-compatible way. Dependent Haskell supports general recursion
without termination checks, which makes it less suitable for theorem
proving. While we share similar goals, our work is differentiated by the
contrasting paradigms of Scala and Haskell. Like many object-oriented
languages, Scala is primarily built around subtyping and does not
restrict the use of side effects. Furthermore, Eisenberg's system
provides control over the relevance of values and type parameters. In
contrast, our system does not support any erasure annotations and simply
follows Scala's canonical erasure strategy: types are systematically
erased to JVM types, and terms are left untouched. Weirich established a
fully mechanized type safety proof for the core of Dependent Haskell
using the Coq proof assistant \autocite{dependentHaskellSpec}.

Cayenne is a Haskell-like language with dependent types introduced in
1998 by Augustsson \autocite{cayenne}. Like Dependent Haskell, it
resembles our system in its treatment of termination, and differs by
being a purely functional programming language. Cayenne's treatment of
erasure is similar to Scala's: types are systematically erased.
Augustsson proves that Cayenne's erasure is semantics-preserving, but
does not provide any other metatheoretical results.

Adding dependent types to object-oriented languages is a remarkably
under-explored area of research. A notable exception is the recent work
of Kazerounian et al.~on adding dependent types to Ruby
\autocite{kazerounian2019type}. Their goals are very much aligned with
ours: using type-level programming to increase program safety. Given the
extremely dynamic nature of Ruby, it is unsurprising that their solution
greatly differs from ours. In their work, type checking happens entirely
at runtime and has to be performed at every function invocation to
account for possible changes in function definitions. Safety is obtained
by inserting dynamic checks, similarly to gradual typing.

The work of Campos and Vasconcelos on DOL (Dependent Object-oriented
Language) \autocite{campos2018dependent} shares similar goals but is
limited to inequality constraints on integer parameters (in the style of
\autocite{xi1998eliminating}).

Dependently-typed lambda calculi with subtyping were described at least
as far back as 1988 by Cardelli \autocite{cardelli1988structural}. His
type system is much more expressive than ours and allows bounded
quantification over both types and terms using the notion of a
\lstinline!Type! type and power types. Unlike our system, which is
designed with the concrete evaluation of types in mind, Cardelli does
not provide semantics for his system and leaves the equivalence relation
among types unspecified.

In \autocite{aspinall1994singleton} Aspinall introduces \({\lambda}_{\leq}{}\), a
dependently-typed system with subtyping \emph{and} singleton types that
resembles ours in its type language. His equivalence relation on types
is more powerful and is not syntax-directed, unlike our type evaluation
relation. Furthermore, singleton types in his work are indexed by the
type through which equality is ``viewed'', thereby enabling a form of
polymorphism beyond ours. Aspinall's system also has primitive types and
allows for atomic subtyping among them, but no congruence rules, hence
partially-widened forms like \(\singleton{ \cons~\choos[\Any]~\nil }\)
cannot be represented.

System \({\lambda}P_{\leq}\) \autocite{aspinall1996dependent} combines subtyping and
dependent types in the Edinburgh Logical Framework. In this work,
Aspinall et al.~propose a type-checking algorithm for \({\lambda}P_{\leq}\) which
they show to be complete and terminating. Their system uses a kinding
relation to ensure well-formedness of type applications. A kind system
is not required in \oursystem as we emulate type applications inside
singleton types.

In \autocite{stone2000deciding}, Stone and Harper describe a
dependently-typed calculus with singleton kinds and subkinding. Their
type-and-kind system is similar to Aspinall's \({\lambda}_{\leq}{}\) term-and-type
system, but operates one level up the hierarchy.

More recently, Courant \autocite{courant2003strong} developed a variant
of Aspinall's \({\lambda}_{\leq}{}\) with a type-inference algorithm that he proves
to be sound and complete. The main takeaway from Courant's work is the
inclusion of a coercion rule in delta reduction. These coercions are
used to ``tag'' variables with their declared type, which prevents these
types from being lost during substitution. Our formalism resembled
Courant's system, it shares the \textsc{SubSing} subtyping rule
(SUB/SINGL in Courant's work), and \(\beta\delta\)-reduction.

Pure Type Systems \autocite{barendregt1991introduction} provide a
unified presentation of systems of dependently-typed
\(\lambda\)-calculus by using a single syntactic category for both terms
and types.

In \autocite{zwanenburg1999pure}, Zwanenburg defines an extension of
pure type systems that include both subtyping and bounded
quantification. A central design decision of his system is that
subtyping rules do not depend on typing rules. The absence of
circularity simplifies both the theory and the metatheory, at the cost
of having to define subtyping on pseudoterms rather than only well-typed
terms. Another limitation of Zwanenburg's theory is that it cannot be
extended with a Top-type.

Pure Subtype Systems \autocite{pureSubtypeSystems} is another framework
with unified syntax; it differs from traditional approaches in that it
uses a single relation, subtyping, that subsumes typing, subtyping, and
type evaluation as found in our system. Their system allows for
partially-widened types similar to ours and also enables the computation
with different levels of precision. For instance, it can conclude that
Int + 5 can be approximated as Int. The paper presents a partial
investigation of the metatheory, but the proof of soundness remains
incomplete. Nevertheless, Hutchins reports that he has not been able to
construct a counter-example, even with the addition of fixpoints.

In \autocite{yang2017unifying}, Yang and Oliveira propose a
dependently-typed generalization of System \(F_{\leq}\) with unified syntax
and a single relation that subsumes typing and subtyping. In their
system, type computations are driven by cast operators: each reduction
or expansion step requires an annotation to explicitly instruct the type
checker to take a step. Explicit casts make it possible to allow general
recursion without compromising decidability of type checking. It would
be interesting to study variants of \oursystem based on explicit casts
instead of our finitized fix.

Dependent-object types \autocite{dotPaper} model the core of Scala's
type system and include type members and path-dependent types, which are
not represented in our formalism. Even though they introduce a form of
dependency, path-dependent types were not designed for type-level
computation, rendering their original goals largely orthogonal to ours.

%% file: benchmarks/concat-graph.tex
\begingroup
  \makeatletter
  \providecommand\color[2][]{%
    \GenericError{(gnuplot) \space\space\space\@spaces}{%
      Package color not loaded in conjunction with
      terminal option `colourtext'%
    }{See the gnuplot documentation for explanation.%
    }{Either use 'blacktext' in gnuplot or load the package
      color.sty in LaTeX.}%
    \renewcommand\color[2][]{}%
  }%
  \providecommand\includegraphics[2][]{%
    \GenericError{(gnuplot) \space\space\space\@spaces}{%
      Package graphicx or graphics not loaded%
    }{See the gnuplot documentation for explanation.%
    }{The gnuplot epslatex terminal needs graphicx.sty or graphics.sty.}%
    \renewcommand\includegraphics[2][]{}%
  }%
  \providecommand\rotatebox[2]{#2}%
  \@ifundefined{ifGPcolor}{%
    \newif\ifGPcolor
    \GPcolorfalse
  }{}%
  \@ifundefined{ifGPblacktext}{%
    \newif\ifGPblacktext
    \GPblacktexttrue
  }{}%
  \let\gplgaddtomacro\g@addto@macro
  \gdef\gplbacktext{}%
  \gdef\gplfronttext{}%
  \makeatother
  \ifGPblacktext
    \def\colorrgb#1{}%
    \def\colorgray#1{}%
  \else
    \ifGPcolor
      \def\colorrgb#1{\color[rgb]{#1}}%
      \def\colorgray#1{\color[gray]{#1}}%
      \expandafter\def\csname LTw\endcsname{\color{white}}%
      \expandafter\def\csname LTb\endcsname{\color{black}}%
      \expandafter\def\csname LTa\endcsname{\color{black}}%
      \expandafter\def\csname LT0\endcsname{\color[rgb]{1,0,0}}%
      \expandafter\def\csname LT1\endcsname{\color[rgb]{0,1,0}}%
      \expandafter\def\csname LT2\endcsname{\color[rgb]{0,0,1}}%
      \expandafter\def\csname LT3\endcsname{\color[rgb]{1,0,1}}%
      \expandafter\def\csname LT4\endcsname{\color[rgb]{0,1,1}}%
      \expandafter\def\csname LT5\endcsname{\color[rgb]{1,1,0}}%
      \expandafter\def\csname LT6\endcsname{\color[rgb]{0,0,0}}%
      \expandafter\def\csname LT7\endcsname{\color[rgb]{1,0.3,0}}%
      \expandafter\def\csname LT8\endcsname{\color[rgb]{0.5,0.5,0.5}}%
    \else
      \def\colorrgb#1{\color{black}}%
      \def\colorgray#1{\color[gray]{#1}}%
      \expandafter\def\csname LTw\endcsname{\color{white}}%
      \expandafter\def\csname LTb\endcsname{\color{black}}%
      \expandafter\def\csname LTa\endcsname{\color{black}}%
      \expandafter\def\csname LT0\endcsname{\color{black}}%
      \expandafter\def\csname LT1\endcsname{\color{black}}%
      \expandafter\def\csname LT2\endcsname{\color{black}}%
      \expandafter\def\csname LT3\endcsname{\color{black}}%
      \expandafter\def\csname LT4\endcsname{\color{black}}%
      \expandafter\def\csname LT5\endcsname{\color{black}}%
      \expandafter\def\csname LT6\endcsname{\color{black}}%
      \expandafter\def\csname LT7\endcsname{\color{black}}%
      \expandafter\def\csname LT8\endcsname{\color{black}}%
    \fi
  \fi
    \setlength{\unitlength}{0.0500bp}%
    \ifx\gptboxheight\undefined%
      \newlength{\gptboxheight}%
      \newlength{\gptboxwidth}%
      \newsavebox{\gptboxtext}%
    \fi%
    \setlength{\fboxrule}{0.5pt}%
    \setlength{\fboxsep}{1pt}%
\begin{picture}(4680.00,3276.00)%
    \gplgaddtomacro\gplbacktext{%
      \csname LTb\endcsname
      \put(721,751){\makebox(0,0)[r]{\strut{}$\sfrac{1}{16}$}}%
      \csname LTb\endcsname
      \put(721,943){\makebox(0,0)[r]{\strut{}$\sfrac{1}{8}$}}%
      \csname LTb\endcsname
      \put(721,1135){\makebox(0,0)[r]{\strut{}$\sfrac{1}{4}$}}%
      \csname LTb\endcsname
      \put(721,1327){\makebox(0,0)[r]{\strut{}$\sfrac{1}{2}$}}%
      \csname LTb\endcsname
      \put(721,1519){\makebox(0,0)[r]{\strut{}1}}%
      \csname LTb\endcsname
      \put(721,1711){\makebox(0,0)[r]{\strut{}2}}%
      \csname LTb\endcsname
      \put(721,1903){\makebox(0,0)[r]{\strut{}4}}%
      \csname LTb\endcsname
      \put(721,2095){\makebox(0,0)[r]{\strut{}8}}%
      \csname LTb\endcsname
      \put(721,2287){\makebox(0,0)[r]{\strut{}16}}%
      \csname LTb\endcsname
      \put(721,2479){\makebox(0,0)[r]{\strut{}32}}%
      \csname LTb\endcsname
      \put(721,2671){\makebox(0,0)[r]{\strut{}64}}%
      \csname LTb\endcsname
      \put(721,2863){\makebox(0,0)[r]{\strut{}128}}%
      \csname LTb\endcsname
      \put(721,3055){\makebox(0,0)[r]{\strut{}256}}%
      \csname LTb\endcsname
      \put(900,484){\makebox(0,0){\strut{}0}}%
      \csname LTb\endcsname
      \put(1260,484){\makebox(0,0){\strut{}}}%
      \csname LTb\endcsname
      \put(1620,484){\makebox(0,0){\strut{}100}}%
      \csname LTb\endcsname
      \put(1980,484){\makebox(0,0){\strut{}}}%
      \csname LTb\endcsname
      \put(2340,484){\makebox(0,0){\strut{}200}}%
      \csname LTb\endcsname
      \put(2699,484){\makebox(0,0){\strut{}}}%
      \csname LTb\endcsname
      \put(3059,484){\makebox(0,0){\strut{}300}}%
      \csname LTb\endcsname
      \put(3419,484){\makebox(0,0){\strut{}}}%
      \csname LTb\endcsname
      \put(3779,484){\makebox(0,0){\strut{}400}}%
    }%
    \gplgaddtomacro\gplfronttext{%
      \csname LTb\endcsname
      \put(171,1903){\rotatebox{-270}{\makebox(0,0){\strut{}Compilation time (seconds)}}}%
      \put(2339,154){\makebox(0,0){\strut{}List size}}%
      \csname LTb\endcsname
      \put(2668,2882){\makebox(0,0)[r]{\strut{}implicit concat}}%
      \csname LTb\endcsname
      \put(2668,2662){\makebox(0,0)[r]{\strut{}dependent concat}}%
    }%
    \gplbacktext
    \put(0,0){\includegraphics{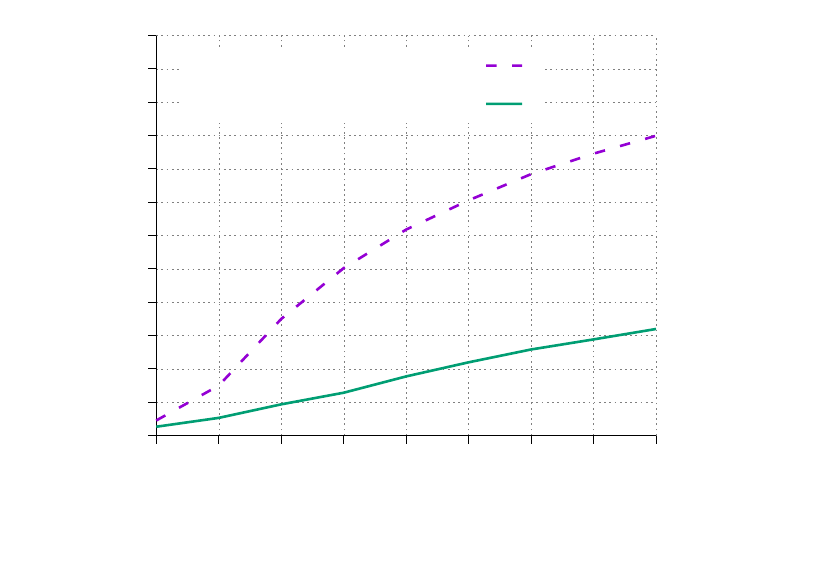}}%
    \gplfronttext
  \end{picture}%
\endgroup

%% file: benchmarks/join-graph.tex
\begingroup
  \makeatletter
  \providecommand\color[2][]{%
    \GenericError{(gnuplot) \space\space\space\@spaces}{%
      Package color not loaded in conjunction with
      terminal option `colourtext'%
    }{See the gnuplot documentation for explanation.%
    }{Either use 'blacktext' in gnuplot or load the package
      color.sty in LaTeX.}%
    \renewcommand\color[2][]{}%
  }%
  \providecommand\includegraphics[2][]{%
    \GenericError{(gnuplot) \space\space\space\@spaces}{%
      Package graphicx or graphics not loaded%
    }{See the gnuplot documentation for explanation.%
    }{The gnuplot epslatex terminal needs graphicx.sty or graphics.sty.}%
    \renewcommand\includegraphics[2][]{}%
  }%
  \providecommand\rotatebox[2]{#2}%
  \@ifundefined{ifGPcolor}{%
    \newif\ifGPcolor
    \GPcolortrue
  }{}%
  \@ifundefined{ifGPblacktext}{%
    \newif\ifGPblacktext
    \GPblacktexttrue
  }{}%
  \let\gplgaddtomacro\g@addto@macro
  \gdef\gplbacktext{}%
  \gdef\gplfronttext{}%
  \makeatother
  \ifGPblacktext
    \def\colorrgb#1{}%
    \def\colorgray#1{}%
  \else
    \ifGPcolor
      \def\colorrgb#1{\color[rgb]{#1}}%
      \def\colorgray#1{\color[gray]{#1}}%
      \expandafter\def\csname LTw\endcsname{\color{white}}%
      \expandafter\def\csname LTb\endcsname{\color{black}}%
      \expandafter\def\csname LTa\endcsname{\color{black}}%
      \expandafter\def\csname LT0\endcsname{\color[rgb]{1,0,0}}%
      \expandafter\def\csname LT1\endcsname{\color[rgb]{0,1,0}}%
      \expandafter\def\csname LT2\endcsname{\color[rgb]{0,0,1}}%
      \expandafter\def\csname LT3\endcsname{\color[rgb]{1,0,1}}%
      \expandafter\def\csname LT4\endcsname{\color[rgb]{0,1,1}}%
      \expandafter\def\csname LT5\endcsname{\color[rgb]{1,1,0}}%
      \expandafter\def\csname LT6\endcsname{\color[rgb]{0,0,0}}%
      \expandafter\def\csname LT7\endcsname{\color[rgb]{1,0.3,0}}%
      \expandafter\def\csname LT8\endcsname{\color[rgb]{0.5,0.5,0.5}}%
    \else
      \def\colorrgb#1{\color{black}}%
      \def\colorgray#1{\color[gray]{#1}}%
      \expandafter\def\csname LTw\endcsname{\color{white}}%
      \expandafter\def\csname LTb\endcsname{\color{black}}%
      \expandafter\def\csname LTa\endcsname{\color{black}}%
      \expandafter\def\csname LT0\endcsname{\color{black}}%
      \expandafter\def\csname LT1\endcsname{\color{black}}%
      \expandafter\def\csname LT2\endcsname{\color{black}}%
      \expandafter\def\csname LT3\endcsname{\color{black}}%
      \expandafter\def\csname LT4\endcsname{\color{black}}%
      \expandafter\def\csname LT5\endcsname{\color{black}}%
      \expandafter\def\csname LT6\endcsname{\color{black}}%
      \expandafter\def\csname LT7\endcsname{\color{black}}%
      \expandafter\def\csname LT8\endcsname{\color{black}}%
    \fi
  \fi
    \setlength{\unitlength}{0.0500bp}%
    \ifx\gptboxheight\undefined%
      \newlength{\gptboxheight}%
      \newlength{\gptboxwidth}%
      \newsavebox{\gptboxtext}%
    \fi%
    \setlength{\fboxrule}{0.5pt}%
    \setlength{\fboxsep}{1pt}%
\begin{picture}(4680.00,3276.00)%
    \gplgaddtomacro\gplbacktext{%
      \csname LTb\endcsname
      \put(721,751){\makebox(0,0)[r]{\strut{}$\sfrac{1}{16}$}}%
      \csname LTb\endcsname
      \put(721,943){\makebox(0,0)[r]{\strut{}$\sfrac{1}{8}$}}%
      \csname LTb\endcsname
      \put(721,1135){\makebox(0,0)[r]{\strut{}$\sfrac{1}{4}$}}%
      \csname LTb\endcsname
      \put(721,1327){\makebox(0,0)[r]{\strut{}$\sfrac{1}{2}$}}%
      \csname LTb\endcsname
      \put(721,1519){\makebox(0,0)[r]{\strut{}1}}%
      \csname LTb\endcsname
      \put(721,1711){\makebox(0,0)[r]{\strut{}2}}%
      \csname LTb\endcsname
      \put(721,1903){\makebox(0,0)[r]{\strut{}4}}%
      \csname LTb\endcsname
      \put(721,2095){\makebox(0,0)[r]{\strut{}8}}%
      \csname LTb\endcsname
      \put(721,2287){\makebox(0,0)[r]{\strut{}16}}%
      \csname LTb\endcsname
      \put(721,2479){\makebox(0,0)[r]{\strut{}32}}%
      \csname LTb\endcsname
      \put(721,2671){\makebox(0,0)[r]{\strut{}64}}%
      \csname LTb\endcsname
      \put(721,2863){\makebox(0,0)[r]{\strut{}128}}%
      \csname LTb\endcsname
      \put(721,3055){\makebox(0,0)[r]{\strut{}256}}%
      \csname LTb\endcsname
      \put(900,484){\makebox(0,0){\strut{}0}}%
      \csname LTb\endcsname
      \put(1260,484){\makebox(0,0){\strut{}}}%
      \csname LTb\endcsname
      \put(1620,484){\makebox(0,0){\strut{}100}}%
      \csname LTb\endcsname
      \put(1980,484){\makebox(0,0){\strut{}}}%
      \csname LTb\endcsname
      \put(2340,484){\makebox(0,0){\strut{}200}}%
      \csname LTb\endcsname
      \put(2699,484){\makebox(0,0){\strut{}}}%
      \csname LTb\endcsname
      \put(3059,484){\makebox(0,0){\strut{}300}}%
      \csname LTb\endcsname
      \put(3419,484){\makebox(0,0){\strut{}}}%
      \csname LTb\endcsname
      \put(3779,484){\makebox(0,0){\strut{}400}}%
    }%
    \gplgaddtomacro\gplfronttext{%
      \csname LTb\endcsname
      \put(391,1903){\rotatebox{-270}{\makebox(0,0){\strut{}}}}%
      \put(2339,154){\makebox(0,0){\strut{}Number of columns}}%
      \csname LTb\endcsname
      \put(3286,1118){\makebox(0,0)[r]{\strut{}implicit join}}%
      \csname LTb\endcsname
      \put(3286,898){\makebox(0,0)[r]{\strut{}dependent join}}%
    }%
    \gplbacktext
    \put(0,0){\includegraphics{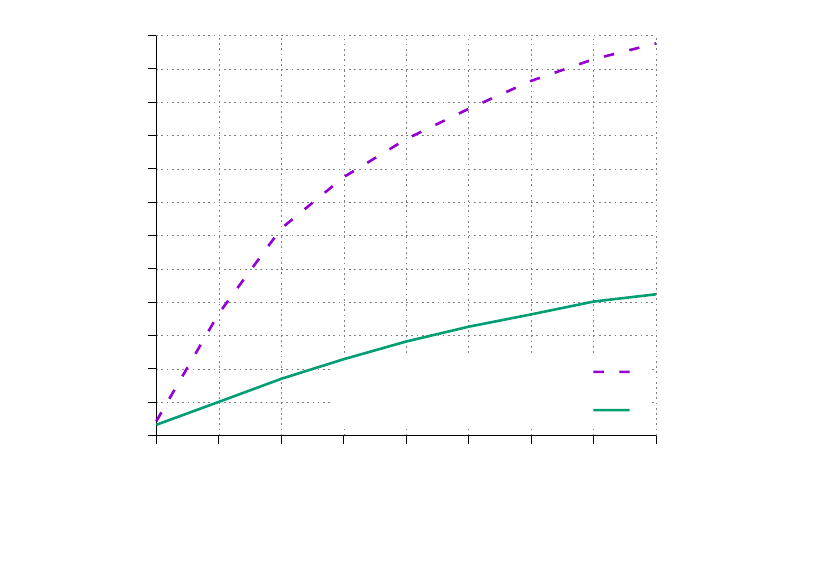}}%
    \gplfronttext
  \end{picture}%
\endgroup